\documentclass[%
 reprint,
superscriptaddress,
 amsmath,amssymb,
 aps,
 twocolumn,
float]{revtex4-1}

\usepackage{graphicx}
\usepackage{dcolumn}
\usepackage{bm}
\usepackage{xcolor}
\usepackage[caption=false]{subfig}
\allowdisplaybreaks

\begin{document}
\title{
Driven One-Particle Quantum Cyclotron 
}

\author{X. Fan}
 \email{xingfan@g.harvard.edu}
\affiliation{Department of Physics, Harvard University, Cambridge, Massachusetts 02138, USA}
 \affiliation{Center for Fundamental Physics, Northwestern University, Evanston, Illinois 60208, USA}
\author{G. Gabrielse}
 \email{gerald.gabrielse@northwestern.edu}
 \affiliation{Center for Fundamental Physics, Northwestern University, Evanston, Illinois 60208, USA}
\date{\today}

\begin{abstract}
A quantum cyclotron is one trapped electron or positron that  occupies only its lowest cyclotron and spin states. A master equation is solved for a driven quantum cyclotron with a QND (quantum nondemolition) coupling to a detection oscillator in thermal equilibrium - the first quantum calculation for this coupled and open system. 
The predicted rate of cyclotron and spin quantum jumps as a function of drive frequency, for a small coupling between the detection motion and its thermal reservoir, differs sharply from what has been predicted and used for past measurements.  The calculation suggests a ten times more precise electron magnetic moment measurement is possible, as needed to investigate current differences between the most precise prediction of the standard model of particle physics, and the most accurate measurement of a property of an elementary particle.  
\end{abstract}

\maketitle

\section{Motivation and Overview}

\begin{figure}[htbp!]
    \centering
    \includegraphics[width=\the\columnwidth]{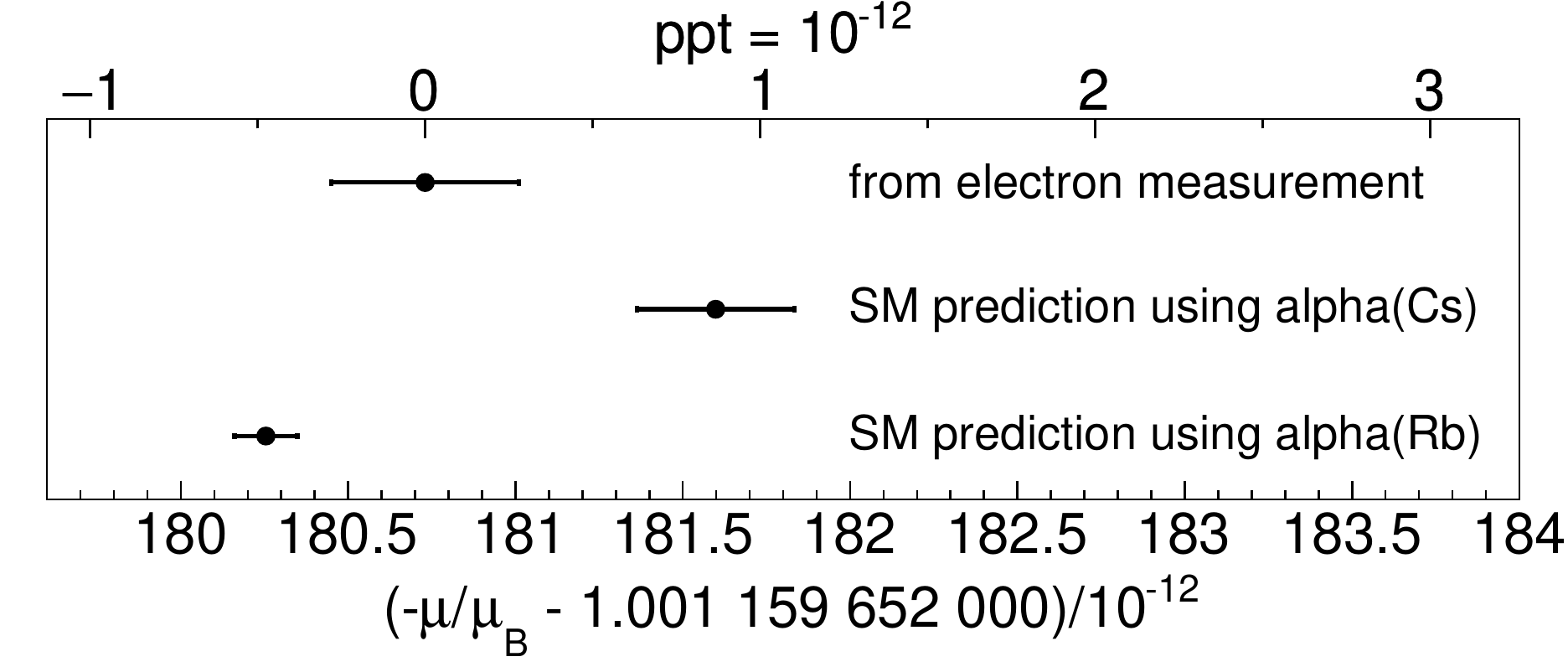}
    \caption{Comparison of the measured electron magnetic moment \cite{HarvardMagneticMoment2008} with the standard model predictions \cite{atomsTheoryReview2019,MullerAlpha2018,RbAlpha2020Nature}} \label{fig:MeasuredAndPredicted}
\end{figure}

An intriguing 2.4 standard deviation discrepancy \cite{atomsNewMeasurement2019,atomsTheoryReview2019,HarvardMagneticMoment2008,MullerAlpha2018} recently arose between the Standard Model's most precise prediction and the measured value (Fig.~\ref{fig:MeasuredAndPredicted}).  The best measurement \cite{HarvardMagneticMoment2008,HarvardMagneticMoment2011} determines the electron magnetic moment in Bohr magnetons ($\mu$/$\mu_B$) to $3$ parts in $10^{13}$ -- the most precisely determined property of an elementary particle.  The SM prediction  requires Dirac theory, quantum electrodynamics, hadronic and weak interaction contributions \cite{atomsTheoryReview2019}. 
The part in $10^{12}$ agreement  between SM prediction and measurement that stood for years gave way as a result of a more precise measurement of the latter. The discrepancy triggered new theoretical investigations into possible physics beyond the SM  \cite{gardner2019light,ALightComplexScalarForTheElectronAndMuonAnomalousMagneticMoments,PhysRevD.98.075011,PhysRevD.98.113002,PhysRevD.99.095034}. As this work was being reported, a second new  $\alpha$ measurement \cite{RbAlpha2020Nature} contradicted the first, giving a SM prediction that disagrees with electron's measurement by 1.6 standard deviations, but in the other direction.   

A one-particle, quantum cyclotron is at the heart of past and future measurements \cite{HarvardMagneticMoment2008,atomsNewMeasurement2019}. A single electron, suspended indefinitely in a Penning trap, is cooled enough that it initially occupies only one of the two stable cyclotron ground states, one with spin down and one with spin up (Fig.~\ref{fig:quantumstates}).  Transitions are driven between these states and a third, the first excited cyclotron state with spin down.  The state of the quantum cyclotron is detected after the drives are turned off using quantum jump spectroscopy.  The angular cyclotron and anomaly drive frequencies, $\omega_c$ and $\omega_a$, that produce one-quantum transitions,  determine the magnetic moment in Bohr magnetons, 
\begin{equation}
\pm\frac{\mu}{\mu_B} = 1 +  \frac{\omega_a}{\omega_c} = \frac{g_\pm}{2}. 
\end{equation}
The plus and minus signs are for the positron and electron, and the g-values $g_\pm$,  divided by 2, are other names for the ratio of moments. The frequency $\omega_c$ is the electron cyclotron frequency.  The anomaly frequency $\omega_a = \omega_s - \omega_c$ is the difference between the electron spin precession frequency $\omega_s$  and its cyclotron frequency. The anomaly frequency $\omega_a$ is directly measured instead of the spin frequency $\omega_s$ because the uncertainty in $\mu_/\mu_B$ is thereby reduced by about a factor of $\omega_c/\omega_a\approx10^3$ \cite{Review}. The resonance line shapes from which these frequencies are extracted have intrinsically different shapes.  

The use of quantum nondemolition (QND) detection methods completely evades detection backaction for determining the quantum state of the cyclotron and spin motion. Nonetheless, detection backaction  still prevented  better measurements of the cyclotron and spin transition frequencies to better determine the  magnetic moments. This backaction produced a very wide and asymmetric quantum jump spectroscopy line shape when cyclotron transitions were driven to determine the cyclotron frequency.  Even though resonant frequencies can be extracted from broad and asymmetric lines in principle, in practice this causes a susceptibility to systematic uncertainties. Significant progress in precision frequency measurements typically takes place only when narrower and more symmetric line shapes are produced. We recently proposed a very promising method for circumventing this detector backaction for the frequency measurements \cite{Fan2020BackActionPRL}.  The cyclotron line shape would be much more symmetric, and orders of magnitude narrower, than for previous measurements.  

In this work we describe the quantum calculation that is carried out to predict the narrow quantum-jump line shapes \cite{Fan2020BackActionPRL} much more completely. A master equation is solved for a driven quantum cyclotron with a QND coupling to a detection oscillator, the latter being coupled to a thermal reservoir.  The predicted quantum jump line shapes are very different than was predicted for the case when the detection oscillator was more strongly coupled to its environment  \cite{BrownLineshapePRL,BrownLineshape}.  We also present for the first time quantum calculations for (1) driven anomaly transitions, (2) directly driven spin flips, and (3) spin flips produced by simultaneous cyclotron and anomaly drives.  The additional calculations make it possible to evaluate and contrast possibilities for making new measurements of the electron and positron magnetic moments.  

Key to these calculations, and the possibility to measure the electron and positron magnetic moments much more accurately, is decoupling  the detection oscillator from its thermal environment by a factor of 100 during the time in the measurement when one-quantum transitions are being driven.    The parameters used in the calculation are those realized in a very recent experimental demonstration of one way that this could be done \cite{FanRFSwitch2020}, while also allowing the necessary coupling to be restored for quantum state readout. 

The outcome of the calculation is that it now seems feasible to carry out new electron and positron magnetic measurements that are an order of magnitude more accurate than was previously possible.  This would make it possible to investigate the discrepancies between the most precise prediction of the standard model of particle physics, and the most accurate measurement of a property of an elementary particle \cite{atomsNewMeasurement2019,atomsTheoryReview2019}. 

Details of the quantum system are given in Sec.~\ref{sec:QuantumCyclotron}. The Hamiltonian and master equation of the system are presented in Sec.~\ref{sec:HamiltonianForTheQuantumSystem}. Calculations of single photon excitations of cyclotron and anomaly transitions are given in Sections~\ref{sec:OneDriveExcitations} and \ref{sec:AnomalyLineshape}, respectively.
Sec.~\ref{sec:DirectSpinFlips} does the same for directly driven spin flips. Sec.~\ref{sec:twophotoncalculation} predicts the quantum-jump line shape for simultaneously applied cyclotron and anomaly drives. Sec.~\ref{sec:Comparison} contrasts the relative advantages of the different methods, and Sec.~\ref{sec:summary} provides a summary.

\section{Quantum Cyclotron}
\label{sec:QuantumCyclotron}

A one-electron quantum cyclotron is at the heart of the approach being investigated here.  An electron or positron in a Penning trap is confined within a spatially uniform magnetic field  $B\hat{z}$, along with an electrostatic quadrupole potential \cite{Review}.  The possibility to use only the ground and first excited cyclotron states of a single isolated electron has already been demonstrated and used for measurement \cite{HarvardMagneticMoment2008}.  The two lowest levels of the quantum cyclotron are separated by an energy $\hbar \omega_c$, where $\omega_c$ is the angular cyclotron frequency introduced above.  The spin up (quantum number $m_s=1/2$) and spin down ($m_s=-1/2$) states are separated in energy by $\hbar \omega_s$, where $\omega_s$ is the spin precession frequency discussed above. This one-particle quantum cyclotron has a Hamiltonian 
\begin{equation}
H=\hbar \omega_s \left(a_s^\dagger a_s 
- \tfrac{1}{2} 
\right)  +\hbar \omega_c  (a_c^\dagger a_c + \tfrac{1}{2}). 
\label{eq:HQuantumCyclotron}
\end{equation}
The spin raising and lowering operators are
\begin{equation}
 \begin{split}
 a_s^\dagger\left|\downarrow\right\rangle&=\left|\uparrow\right\rangle \\ a_s\left|\uparrow\right\rangle&=\left|\downarrow\right\rangle,  
 \label{eq:spinoperator}
\end{split}
\end{equation}
 and $a_c^\dagger$ and $a_c$ are harmonic raising and lowering operators for the cyclotron motion \cite{Review}. 

An electrostatic quadrupole potential added to the magnetic field makes a Penning trap that can suspend a single charged particle indefinitely within an extremely high vacuum \cite{PbarMass}.  The electron (of charge $-e$ and mass $m$) oscillates along the magnetic field direction in a  harmonic oscillator potential energy,
\begin{equation}
W(z)=\tfrac{1}{2} m \omega_z^2 z^2
\end{equation}
and $\omega_z$ is the angular axial oscillation frequency. The electrostatic quadrupole shifts the cyclotron frequency slightly in a well understood way \cite{InvarianceTheorem,Review} that can be neglected for the purposes of this calculation.  

This axial motion is used to make quantum nondemolition (QND) measurements of one-quantum spin and cyclotron transitions \cite{QuantumCyclotron,QNDScience1980,QNDReview1980, QNDreview1996,1996MarkQND}. A small magnetic bottle gradient, $B_2 z^2$, is added to the spatially uniform magnetic field, $B_0$, of the Penning trap,  
The addition modifies the axial trapping potential and shifts the frequency of the axial oscillation. A QND detection of a one-quantum cyclotron excitation is possible because it shifts 
the axial frequency from $\omega_z$ to  $\omega_z+\delta_c$, with 
\begin{equation}
    \delta_c = \frac{e B_2}{m} \frac{\hbar}{m \omega_z} \approx 2\pi~ \times(3~\rm{Hz})
\end{equation}\cite{Review}, without changing the cyclotron state. (A two-quantum cyclotron excitation would be $2 \delta_c$ and so on, as will be discussed later and quantified in  Eq.~(\ref{eq:axialfrequencyshift}). The one-quantum shift is just large enough to be detectable.  The relative shift is $\delta_c/\omega_z = 1.5 \times 10^{-8}$ for demonstrated experimental values \cite{HarvardMagneticMoment2011} ($B_2 = 1500$ T/m$^2$ and $\omega_z/(2 \pi ) = 200$ MHz.). 
This bottle shift can be decreased in two ways -- by decreasing the magnetic gradient $B_2$ or by increasing the axial frequency, $\omega_z$.  
Since a next generation  experiment\cite{atomsNewMeasurement2019} uses $B_2=660$ T/m$^2$, we choose the intermediate value $B_2=1200$ T/m$^2$ for the illustrations in this paper.  

The magnetic gradient is unfortunately also responsible for a backaction that broadens the range of frequencies over which a driven cyclotron excitation or spin flip can occur.  The cyclotron and spin frequencies in Eq.~(\ref{eq:HQuantumCyclotron}) both acquire a small $z^2$ dependence,  
\begin{eqnarray}
&&\omega_{c}(z)=\omega_{c} + \tfrac{eB_2}{m}z^2\\ &&\omega_{s}(z)=\omega_{s} + \tfrac{g}{2}\tfrac{eB_2}{m}z^2,
\label{eq:spindependentfreq}
\end{eqnarray}
where the g-value is $g_+$ for a positron and $g_-$ for an electron.  A one-quantum axial excitation within the magnetic bottle gradient shifts the cyclotron frequency by the same $\delta_c$.  A thermal distribution over $\bar{n}_z$ axial states (Eq.~(\ref{eq:Averagenz})) thus makes the cyclotron frequency fluctuate  over a spread of frequencies that is of order $\bar{n}_z \delta_c$. 

Two relativistic shifts must be mentioned, both arising from the ``relativistic mass increase."  The largest is the increase of the effective mass due to the energy of cyclotron motion \cite{Review},
\begin{equation}
\delta_{r} = -\frac{\hbar\omega_c}{mc^2} \omega_c  \approx - 2\pi\times (180~\rm{Hz}).    
\end{equation}
It is only a $1$ part in $10^9$ shift of the cyclotron frequency per cyclotron quantum, but it is a large shift compared to the experimental precision that can being attained. The cyclotron transition frequency between quantum number $n_c$ and $n_c+1$ is shifted by $(n_c+\tfrac{1}{2})\,\delta_r$. The cyclotron frequency between the ground and first excited cyclotron states with spin down shift by half of $\delta_r$.   The shift is thus extremely important in that a cyclotron drive that excites the first spin-down excited state, will not excite a cyclotron excitation of the spin-up ground state. 
However, for the purposes of this calculation it can simply be absorbed into $\omega_c$.

The second relativistic shift, 
\begin{equation}
\delta_{cr} = -\frac{\hbar\omega_c}{2mc^2} \omega_z  \approx - 2\pi\times (0.12~\rm{Hz}),    \end{equation}
is about 1000 times smaller.  It 
comes from the increase of the effective mass due to the zero-point energy of the axial oscillation.  This coupling has much the same effect in coupling the motions to allow QND detection as does a magnetic bottle \cite{Review}.  It also produces a corresponding backaction. This relativistic coupling is neglected here because it is 25 times smaller than the coupling caused by the magnetic bottle gradient considered above.

A spin flip shifts the angular axial frequency by $\delta_s = (g/2) \,\delta_c$.  This is nearly the same size as the corresponding cyclotron frequency shift because $\frac{g}{2}$ differs from 1 by only a part in 1000, and experiments are not able to resolve these two shifts from each other. The frequency difference $\omega_a=\omega_s-\omega_c$ is measured rather than $\omega_s$ \cite{HarvardMagneticMoment2008}, and the thousand times smaller shift, $\delta_a = \delta_s-\delta_c$, is thus also important.  

Table~\ref{table:Frequencies} gives the typical  trapped electron frequencies, damping rates, and quantum numbers used in this calculation.  The spin and cyclotron frequencies are for an electron in a $B=5.3$ T magnetic field, and $\gamma_c$ is the rate at which the first excited cyclotron state radiates spontaneous emission to return to its ground state. This radiation rate is substantially inhibited by a surrounding cylindrical trap cavity \cite{CylindricalPenningTrapDemonstrated, InhibitionLetter,HarvardMagneticMoment2011}. The spin-up cyclotron ground state radiates with a time constant so long that we treat it as stable.  

The axial frequency depends upon the trap size and the applied trapping potential \cite{CylindricalPenningTrap,Gabrielse84h}.  Its damping rate $\gamma_z$ depends upon the quality factor and inductive reactance of the damping and detection circuit to which it is coupled \cite{ElectronCalorimeter}. The maximal damping rate in Tab.~\ref{table:Frequencies} applies during particle detection.  For this calculation, we assume that this rate is electronically reduced by a factor of 100 during the time that spin and cyclotron transitions are driven, a number that has been experimentally demonstrated \cite{FanRFSwitch2020}.  The average quantum number is for thermal equilibrium with a circuit kept at 0.1 K, the ambient temperature that has been maintained for measurements  using a dilution refrigerator \cite{HarvardMagneticMoment2008}.  

The magnetron orbit of a trapped particle is important experimentally but not for this calculation.  It is a motion at a much lower frequency.  The average quantum number in the table pertains for the sideband cooling limit \cite{Review}, and its radiation damping rate is completely negligible. The broadening due to magnetron motion is smaller than that due to axial motion by a factor of $\omega_m/\omega_z\approx1/1000$, and we drop the magnetron motion term to simplify the calculation. If necessary, the Hamiltonian and master equation in Sec.~\ref{sec:HamiltonianForTheQuantumSystem} and Sec.~\ref{sec:OneDriveExcitations} can be naturally generalized to include it.

\begin{table}[htbp!]
\begin{tabular}{c|l|l|l}
                & frequency & \begin{tabular}{@{}c@{}}damping\\time\end{tabular}  &   \begin{tabular}{@{}c@{}}quantum\\number\end{tabular}   \\
\hline\hline
spin        & $\omega_s/2\pi\approx148.5$ GHz& $\gamma_s^{-1}\approx 10^8$ s   & $m_s=\pm\tfrac{1}{2}$ \\
cyclotron   & $\omega_c/2\pi\approx148.3$ GHz& $\gamma_c^{-1}\approx 5$ s   & $\bar{n}_c=0$ \\
axial       & $\omega_z/2\pi\approx 200$ MHz& $\gamma_z^{-1}\approx 0.2$ s   & $\bar{n}_z=10$ \\
magnetron   & $\omega_m/2\pi\approx 133$ kHz& $\gamma_m^{-1}\approx10^{17}$ s   & $\bar{n}_m=10$ \\
anomaly   & $\omega_a/2\pi\approx 170$ MHz& ---   & --- 
\end{tabular}
\caption{The frequencies, damping rates, and quantum numbers used for this calculation are typical for an electron in a Penning trap \cite{HarvardMagneticMoment2008}.}
\label{table:Frequencies}
\end{table}

Tables ~\ref{table:Frequencies} and \ref{table:FrequenciesCompared} list the parameters  used for this calculation.  They are mostly what has been realized experimentally.  Table 
\ref{table:Frequencies} gives  frequencies, damping times and quantum number for the spin, cyclotron, axial an magnetron motion of an electron or positron in a Penning trap. Table~ \ref{table:FrequenciesCompared} compares the important frequency offsets and corresponding time constants. 

\begin{table}[htbp!]
\begin{tabular}{c |c |c}
  ang.\ frequency or rate & frequency (Hz) & time constant (s) \\
\hline
\hline
$\delta_a$          & $0.003$  & $60$\\
$\gamma_z$          & $0.003$  & $60$\\
$\bar{n}_z\delta_a$ & $0.03$   & $6$\\
$\gamma_c$          & $0.03$   & 6\\
$\bar{n}_z\gamma_z$ & $0.03$   & $6$\\
$\delta_c$          & $3$      & $0.06$\\
$\bar{n}_z\delta_c$ & $30$     & $0.006$
\label{tab:twodrivefrequencyscale}
\end{tabular}
\caption{Hierarchy of angular frequencies and rates that are in reach for a new generations of measurements.  The numerical values are frequencies in Hz and times in seconds, with $\delta_a/2\pi = 0.003 $ Hz and $\delta_a^{-1} = 60$ s, for example.}
\label{table:FrequenciesCompared}
\end{table}

One motivation for this calculation is evaluating the possibilities that open if a greatly reduced axial damping rate pertains while cyclotron and anomaly transitions are driven.    The rate can be electronically switched \cite{FanRFSwitch2020} to the low value in the table just before drives are applied, to make one-quantum anomaly and cyclotron transitions with an electron largely uncoupled from the bath.    After the drives are turned off, the damping rate can be electronically switched to a much larger values, as needed to detect the particle state and to damp the axial motion.

\section{Hamiltonian}
\label{sec:HamiltonianForTheQuantumSystem}

\newcommand{\Schro}{Schr\"{o}dinger~}

The basic Hamiltonian for the quantum cyclotron,
\begin{equation}
H_0    =\hbar\omega_s\left(a_s^{\dagger}a_s-\tfrac{1}{2}\right) 
+ \hbar\omega_c
\left(a_c^{\dagger}a_c+\tfrac{1}{2}\right) +\hbar \omega_z \left(a_z^{\dagger}a_z+\tfrac{1}{2}\right),
\label{eq:H0}
\end{equation}
is the sum of independent spin, cyclotron and axial terms. 
The raising and lowering operators for the spin ($a_s^\dagger$ and $a_s$), cyclotron ($a_c^\dagger$ and $a_c$) and axial ($a_z^\dagger$ and $a_z$) motions are introduced in Ref.~\cite{Review}, along with  relationships to the position and momentum operators.  The eigenstates for $H_0$ are direct products of independent spin, cyclotron, and axial eigenstates  $\left|m_s,n_c,n_z\right\rangle$, with
\begin{equation}
E_0(m_s,n_c,n_z)
=\hbar\omega_sm_s+\hbar\omega_c\left(n_c+\tfrac{1}{2}\right)+\hbar\omega_z\left(n_z+\tfrac{1}{2}\right)
\label{eq:E0}
\end{equation}
as the resulting energy eigenvalues, with $m_s = \pm 1/2$, $n_c=0, 1, ...$ and $n_z=0,1,...\,$.
  The magnetron motion of a particle in a Penning trap is neglected because it introduces no significant complications, and because it can be cooled to a small radius that does not change during a measurement.

The addition of a magnetic bottle gradient adds a coupling term to make the Hamiltonian, $H = H_0 + V$,  with
\begin{equation}
V        =    \tfrac{\hbar}{2}\left[\delta_s \left(a_s^{\dagger}a_s-\tfrac{1}{2}\right)+\delta_c \left(a_c^{\dagger}a_c+\tfrac{1}{2}\right)  \right] (a_z^\dagger + a_z)^2, \label{eq:V}
\end{equation}
when contributions smaller by order $\omega_z/\omega_c$ are neglected.
This is a QND coupling because  $[H_0,V]=0$.  The result is that the energy eigenstates of $H=H_0+V$ are the same uncoupled states $\left|n_c, m_s, n_z\right\rangle$ that are the energy eigenstates of $H_0$. The magnetic bottle shifts the energy eigenvalues to   
\begin{equation}
\begin{split}
 E(m_s,&n_c,n_z) = E_0(m_s,n_c,n_z) \\
&+\hbar\delta_c\left(n_c+\tfrac{1}{2}\right)\left(n_z+\tfrac{1}{2}\right) + \hbar\delta_sm_s\left(n_z+\tfrac{1}{2}\right).
\label{eq:energystate}
\end{split}
\end{equation}
That this coupling makes it possible to detect that quantum spin and cyclotron states can be seen by rewriting the energy eigenvalues as
\begin{equation}
E(m_s,n_c,n_z)
=\hbar\omega_sm_s+\hbar\omega_c\left(n_c+\tfrac{1}{2}\right)+\hbar\widetilde{\omega}_z\left(n_z+\tfrac{1}{2}\right).
\label{eq:0}
\end{equation}
Monitoring the effective axial oscillation frequency
\begin{equation}
    \widetilde{\omega}_z=\omega_z+m_s \delta_s+(n_c+\tfrac{1}{2})\delta_c, 
    \label{eq:axialfrequencyshift}
\end{equation}
thus reveals the spin and cyclotron states via their quantum numbers.  A feature of the QND detection is that the axial detection backaction upon these quantum states is completely evaded. Repeated measurements, made to see if something else is changing these states, do not in themselves change the quantum state.

Critical to this work is that the QND coupling $V$ that completely evades detection backaction in the determination of the quantum spin and cyclotron states, does not do so for a measurement of either $\omega_s$ or $\omega_c$. This can be seen by writing the energy eigenvalues in the alternate form,   
\begin{equation}
E(m_s,n_c,n_z) =  \hbar \widetilde{\omega}_s m_s+
\hbar \widetilde{\omega}_c (n_c+\tfrac{1}{2}) + \hbar \omega_z (n_z+\tfrac{1}{2}). 
\end{equation}
Despite the QND coupling, the effective spin, cyclotron and anomaly frequencies all have shifts that go as the axial quantum number 
\begin{subequations}\begin{align}
 \widetilde{\omega}_s &= \omega_s + \delta_s (n_z+\tfrac{1}{2}),\\ 
  \widetilde{\omega}_c &= \omega_c + \delta_c (n_z+\tfrac{1}{2}),\\  
  \widetilde{\omega}_a &= \omega_a + \delta_a (n_z+\tfrac{1}{2}).  
 \label{eq:CyclotronShift}
 \end{align}
 \label{eq:Shifts}%
 \end{subequations}
These detection backaction shifts cannot  be completely evaded because a shift due to axial zero point motion remains even if the axial detection motion would be cooled to its $n_z=0$ ground state. Because the shifts in this limit are orders of magnitude smaller than what has been attained, we focus upon how these zero-point limits can be attained.  We call this ``circumventing'' detection backaction because of the possibility to achieve these limits while axial detection states well above $n_z=0$ are populated \cite{Fan2020BackActionPRL}.

Electron and positron magnetic moment measurements require the determination of the cyclotron and the anomaly frequencies, $\omega_c$ and $\omega_a$.  These frequencies can be determined observing the rate of quantum jumps between the lowest cyclotron and spin states as a function of the frequency of external driving forces introduced to make these transitions.  
Because the axial detection motion is coupled to a thermal reservoir there is a thermal distribution of axial states.    This spreads out the range of spin,  cyclotron and anomaly frequencies at which a spin, cyclotron or anomaly drive causes one-quantum transitions. The detection backaction thus significantly broadens the observed spin, cyclotron and anomaly resonance line shapes from which the needed frequencies must be detected.

Switching from the \Schro  picture to the interaction picture transforms away the 
well-understood spin, cyclotron and axial motions in the absence of a magnetic bottle.  Terms that go as $a_za_z$ and $a_z^\dagger a_z^\dagger$ oscillate rapidly and hence average to zero in the interaction picture.  The resulting interaction Hamiltonian $\widetilde{V} = e^{i H_0 t/\hbar} V e^{-i H_0 t/\hbar}$ is   
\begin{equation}
\begin{split}
\widetilde{V}=&\left[
\hbar\delta_s\left(a_s^{\dagger}a_s-\tfrac{1}{2}\right)
+\hbar\delta_c\left(a_c^{\dagger}a_c+\tfrac{1}{2}\right)
\right] \left(a_z^{\dagger}a_z+\tfrac{1}{2}\right).
\end{split}
\end{equation}
We  continue using the time-independent raising and lowering operators from the \Schro picture   (rather than transforming these to the interaction picture).  The interaction picture Hamiltonian has an energy scale set by the tiny bottle shifts, $\delta_c$ and $\delta_s$, rather than by the much larger frequencies $\omega_c$, $\omega_s$ and $\omega_z$. 

\begin{figure}
    \centering
    \includegraphics[width=\the\columnwidth]{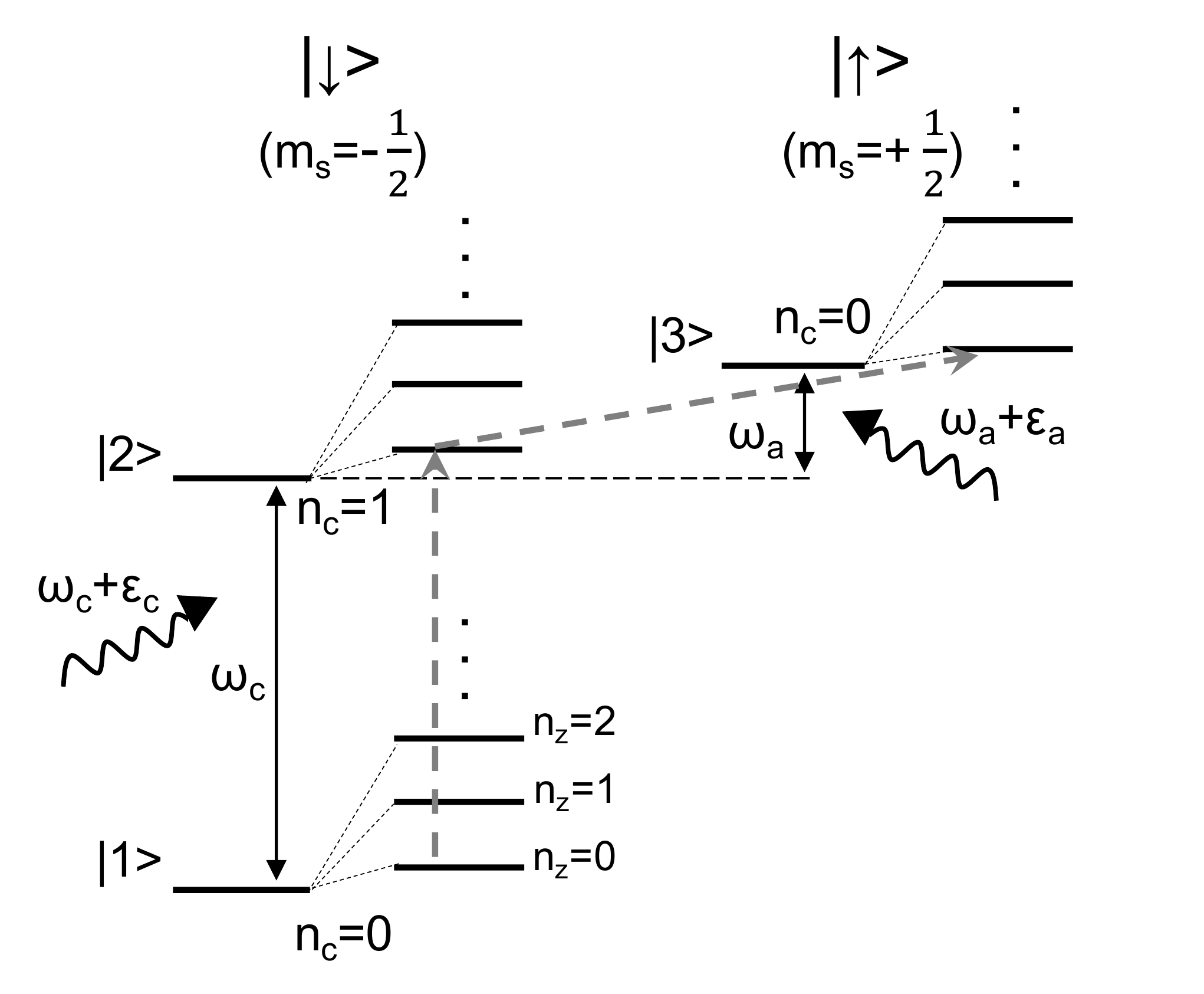}
    \caption{Quantum states of a particle in a Penning Trap. Cyclotron transition and anomaly transition of $n_z=0$ state are also shown with dotted lines. Each cyclotron state has infinite number of axial substates.} \label{fig:quantumstates}
\end{figure}

Figure \ref{fig:quantumstates} represents the lowest of these quantum energy levels, with spin down states ($m_s=-1/2$) on the left and spin up states ($m_s=1/2$) on the right.   The lowest of the infinite ladder of cyclotron states are shown ($n_c = 0, 1$), as are the lowest three of the infinite ladder of axial states ($n_z=0, 1, 2$). For the driving forces we will consider, the electron will essentially occupy only the three cyclotron and spin state combinations
\begin{equation}
\begin{split}
\left|1,n_z\right\rangle&\equiv\left|n_c=0, m_s=-\tfrac{1}{2},n_z\right\rangle,\\ \left|2,n_z\right\rangle&\equiv\left|n_c=1, m_s=-\tfrac{1}{2},n_z\right\rangle,\\ \left|3,n_z\right\rangle&\equiv\left|n_c=0, m_s=+\tfrac{1}{2},n_z\right\rangle,
\end{split}
\label{eq:ThreeStates}
\end{equation}
with $n_z = 0, 1, ...$. 
These are the basis of time-independent states used for this calculation.  The basis would shrink to only three states if the axial motion would be cooled to its quantum ground state.  

The electromagnetic drives that oscillate at angular frequencies $\omega_s + \epsilon_s$,  $\omega_c + \epsilon_c$ and $\omega_a + \epsilon_a$ to drive spin, cyclotron and anomaly transitions are described by the Hamiltonians  
\begin{eqnarray}
V_s(t) &=&  \tfrac{1}{2} \hbar \Omega_s \left[a_s^\dagger e^{-i (\omega_s + \epsilon_s) t} + a_s e^{i (\omega_s + \epsilon_s) t} \right] 
\label{eq:Drives}\\
V_c(t) &=&  \tfrac{1}{2} \hbar \Omega_c \left[a_c^\dagger e^{-i (\omega_c + \epsilon_c) t} + a_c e^{i (\omega_c + \epsilon_c) t} \right] 
\label{eq:Drivec}\\
V_a(t) &=&  \tfrac{1}{2} \hbar \Omega_a \left[a_a^\dagger e^{-i (\omega_a + \epsilon_a) t} + a_a e^{i (\omega_a + \epsilon_a) t} \right].
\label{eq:Drivea}
\end{eqnarray}
 The positive Rabi frequencies $\Omega_s$, $\Omega_c$ and $\Omega_a$ quantify the drive strengths, and $\epsilon_s$, $\epsilon_c$ and $\epsilon_a$ are detunings of the drives from resonance.  
In the interaction picture the Hamiltonian drive terms are
\begin{eqnarray}
\widetilde{V}_s(t)&=&\tfrac{1}{2}\hbar\Omega_s \left[ a^\dagger_s e^{-i\epsilon_s t} + a_s e^{i\epsilon_s t}  \right]\\
\widetilde{V}_c(t)&=&\tfrac{1}{2}\hbar\Omega_c \left[ a^\dagger_c e^{-i\epsilon_c t} + a_c e^{i\epsilon_c t}  \right]\\
\widetilde{V}_a(t)&=&\tfrac{1}{2}\hbar\Omega_a \left[ a^\dagger_a e^{-i\epsilon_a t} + a_a e^{i\epsilon_a t}  \right].
\label{eq:driveeq}
\end{eqnarray}
An anomaly transition is a simultaneous cyclotron and spin transition. The raising operator for an anomaly transition from $\left|2,n_z\right\rangle$ to $\left|3,n_z\right\rangle$, for example, requires  
  $a_a^\dagger=a_s^\dagger a_c$, a lowering of the cyclotron state followed by a raising of the spin state.  A transition from the spin down ground state to the spin up ground state  is accomplished by $a_a^\dagger a_c^\dagger$.

The axial and cyclotron motions are both coupled to a thermal bath, with damping rates of $\gamma_z$ and  $\gamma_c$, respectively. 
An ambient bath temperature of 0.1 K is assumed because it has been demonstrated in experiments \cite{HarvardMagneticMoment2011}. The energy for a one-quantum axial excitation, $\hbar\omega_z/k_B=0.01$ K in temperature units, is instead much smaller than 0.1 K.  The axial state is thus a Boltzmann distribution with an average quantum number 
\begin{equation}
\bar{n}_z=\left[\exp\left({\frac{\hbar\omega_z}{k_BT}}\right)-1\right]^{-1} \approx \frac{k_B T}{\hbar \omega_z}\approx 10.
\label{eq:Averagenz}
\end{equation}
It may be possible to cool this motion further using cavity sideband cooling \cite{atomsNewMeasurement2019}, but this is not assumed here.  A cyclotron excitation requires an energy of $\hbar\omega_c/k_B=7.1$ K that is much larger than the $0.1$ K bath temperature.  The result is that 
 \begin{equation}
     \bar{n}_c=\left[\exp\left({\frac{\hbar\omega_c}{k_BT}}\right)-1\right]^{-1} = 1.2\times10^{-32} \approx 0.
 \end{equation}
 The cyclotron motion  essentially remains in its $n_c=0$ ground state  \cite{QuantumCyclotron} unless an excitation drive is applied.  

For an electron or positron coupled to a thermal bath, a density operator must be used. The density operator in the \Schro picture, $\rho$, and the interaction picture, $\tilde{\rho}$ are related by
\begin{equation}
\widetilde{\rho} = e^{i H_0 t /\hbar} \rho  e^{-i H_0 t/\hbar}. 
\end{equation}
Both $\rho$ and $\widetilde{\rho}$ can be expanded in the infinite base of time-independent states in Eq.~(\ref{eq:ThreeStates}). The diagonal elements are the probabilities to be in each basis state. These are invariant under a change between the \Schro and interaction pictures.  Also invariant are the traces,
\begin{eqnarray}
P_l = \sum_{n_z=0}^{\infty} \langle l,n_z |\rho     \left|l,n_z\right\rangle  = \sum_{n_z=0}^{\infty} \langle l,n_z |\tilde{\rho}     \left|l,n_z\right\rangle,
\label{eq:Pl}
\end{eqnarray}
that are the total probabilities to be in each of the 3 spin and cyclotron states. Here, $l$ denotes the label 1, 2, or 3 introduced in Fig.~\ref{fig:quantumstates} and Eq.~(\ref{eq:ThreeStates}).

The \Schro picture density operator, $\rho$, evolves in time as described by a Lindblad master equation \cite{Lindblad1,Lindblad2,jacobs_2014},
\begin{equation}
\begin{split}
\frac{d\rho}{dt}&=-\frac{i}{\hbar}\left[{H_0} + V + {V}_s + {V}_c + {V}_a,\rho\right] \\
&-\frac{\gamma_c}{2}\left(a_c^\dagger a_c{\rho}-2a_c {\rho} a_c^\dagger+{\rho} a_c^\dagger a_c\right)\\ 
&-\frac{\gamma_z}{2}\bar{n}_z\left(a_za_z^\dagger{\rho}-2a_z^\dagger{\rho} a_z+{\rho} a_za_z^\dagger\right)\\
&-\frac{\gamma_z}{2}\left(\bar{n}_z+1\right)\left(a_z^\dagger a_z{\rho}-2a_z {\rho} a_z^\dagger+{\rho} a_z^\dagger a_z\right).
\end{split}
\label{eq:MasterEquationSchroedinger}
\end{equation}
The coherent time evolution is described by the commutator term.  The incoherent spontaneous emission from the cyclotron motion (from the first excited cyclotron state to its ground state) is described by the nonlinear terms in line two.  (As noted earlier, the heating of the cyclotron motion by the thermal black-body radiation for low temperature surroundings can be neglected.)  The coupling of the axial motion and the thermal bath is described by the last two lines.  The bath temperatures comes in via the average axial quantum number $\bar{n}_z$ of Eq.~(\ref{eq:Averagenz}).

The interaction picture density operator, $\tilde{\rho}$, evolves as
\begin{equation}
\begin{split}
\frac{d\tilde{\rho}}{dt}&=-\frac{i}{\hbar}\left[\widetilde{V}+ \widetilde{V}_s+ \widetilde{V}_c + \widetilde{V}_a,\tilde\rho\right] \\
&-\frac{\gamma_c}{2}\left(a_c^\dagger a_c\tilde{\rho}-2a_c \tilde{\rho} a_c^\dagger+\tilde{\rho} a_c^\dagger a_c\right)\\ 
&-\frac{\gamma_z}{2}\bar{n}_z\left(a_za_z^\dagger\tilde{\rho}-2a_z^\dagger\tilde{\rho} a_z+\tilde{\rho} a_za_z^\dagger\right)\\
&-\frac{\gamma_z}{2}\left(\bar{n}_z+1\right)\left(a_z^\dagger a_z\tilde{\rho}-2a_z \tilde{\rho} a_z^\dagger+\tilde{\rho} a_z^\dagger a_z\right).
\end{split}
\label{eq:MasterEquationI}
\end{equation}
As for the Hamiltonian, we use the time-independent, raising and lowering operators from the \Schro picture. The damping terms transform to have the same form in both pictures. Explicit calculation are done using the interaction picture because it is simpler.  $H_0$ is removed, 
and $\widetilde{V}_s+\widetilde{V}_c + \widetilde{V}_a$ varies much less rapidly in time than does $V_s+V_c + V_a$.

\section{Driven Cyclotron Excitations}
\label{sec:OneDriveExcitations}

\newcommand{\ArgBoth}{\left(
\tfrac{\epsilon}{\delta},
\tfrac{\gamma_c}{\delta},
\tfrac{\gamma_z}{\delta},
\bar{n}_z
\right)}

\newcommand{\ArgC}{\left(
\tfrac{\epsilon_c}{\delta_c},
\tfrac{\gamma_c}{\delta_c},
\tfrac{\gamma_z}{\delta_c},
\bar{n}_z
\right)}

\newcommand{\ArgA}{\left(
\tfrac{\epsilon_a}{\delta_a},
\tfrac{\gamma_c}{\delta_a},
\tfrac{\gamma_z}{\delta_a},
\bar{n}_z
\right)} 

\newcommand{\Arg}{\left(
\tfrac{\epsilon}{\delta},
\tfrac{\gamma_c}{\delta},
\tfrac{\gamma_z}{\delta},
\bar{n}_z}

\subsection{Cyclotron Master Equation}
\label{sec:CyclotronLineshape}

\newcommand{\ket}[2]{\left|#1,#2\right\rangle} 
\newcommand{\bra}[2]{\left\langle#1,#2\right|} 
\newcommand{\me}[4]{\left\langle#1,#2\right|\tilde{\rho}\left|#3,#4\right\rangle} 

A weak cyclotron drive, $V_c$, excites cyclotron states $\left|2,n_z\right\rangle$ from an initial state that is a thermal distribution of spin down, cyclotron ground states, $\left|1,n_z\right\rangle$.    The drive provides no mechanism to flip the spin, so the states $\left|3,n_z\right\rangle$ are not populated. For a weak drive, $\Omega_c \ll \gamma_c$, the probability of a cyclotron excitation is very small.  We neglect the possibility of a second cyclotron excitation that follows the first, from the excited state $\left|2,n_z \right\rangle$ to a higher state, because this is much smaller still.  The Hermitian density operator for cyclotron excitation,
\begin{eqnarray}
\tilde{\rho} &= \tilde{\rho}_{11} + \tilde{\rho}_{12} + \tilde{\rho}_{21} +\tilde{\rho}_{22}
= \begin{pmatrix} \tilde{\rho}_{11} & \tilde{\rho}_{12} \\ \tilde{\rho}_{21} & \tilde{\rho}_{22} \end{pmatrix}
\end{eqnarray}
is the sum of four operators, each defined by \begin{equation}
\tilde{\rho}_{jk} \equiv \sum_{n_z,n^\prime_z} \ket{j}{n_z} \me{j}{n_z}{k}{n^\prime_z} \bra{k}{n^\prime_z}.
\label{eq:rhojk}
\end{equation}
Since $\tilde{\rho}$ is Hermitian, $\tilde{\rho}_{21}=\tilde{\rho}_{12}^\dagger.$

The initial density operator at time $t=0$ is diagonal with respect to the axial quantum numbers, 
\begin{equation}
\begin{split}
&\left\langle 1,n_z|\tilde{\rho}|1,n_z\right\rangle = p_{n_z}(T) =\\ 
&=\left[1-\exp\left({-\frac{\hbar\left(\omega_z-\frac{1}{2}\delta_a\right)}{k_BT}}\right)\right]\exp\left({-\frac{n_z\hbar\left(\omega_z-\frac{1}{2}\delta_a\right)}{k_BT}}\right)\\
&\approx \left[1-\exp\left({-\frac{\hbar\omega_z}{k_BT}}\right)\right]\exp\left({-\frac{n_z\hbar\omega_z}{k_BT}}\right)
\end{split}
\label{eq:11distribution}
\end{equation}
with Boltzmann factors as its nonzero elements.  The approximation is nearly exact because $\delta_a \ll \omega_z$. In the weak drive limit, we would expect this distribution of initial states to remain essentially unchanged.    

The probability $P_2$ from  Eq.~(\ref{eq:Pl}), that the system is excited by one quantum from its spin-down, cyclotron ground state,
\begin{equation}
P_2 = \sum_{n_z} \me{2}{n_z}{2}{n_z}  = \textrm{Tr}\left[\tilde{\rho}_{22}\right].
\label{eq:Pc}
\end{equation}
is the sum of the probabilities for excitation to any of the states $\ket{2}{n_z}$. 
Either the Schrodinger or interaction picture density operator can be used since their diagonal elements are identical.

Determining $\tilde{\rho}_{22}$ requires solving the master equation 
\begin{equation}
\begin{split}
\frac{d}{dt} &\begin{pmatrix} \tilde{\rho}_{11} & \tilde{\rho}_{12} \\ \tilde{\rho}_{21} & \tilde{\rho}_{22} \end{pmatrix} = -{i}\left(a_z^{\dagger}a_z+\tfrac{1}{2}\right)
\begin{pmatrix} 0 & -\delta_c\tilde{\rho}_{12} \\ \delta_c\tilde{\rho}_{21} & 0 \end{pmatrix}\\
&-i\frac{\Omega_c}{2}
\begin{pmatrix}  
i2\mathrm{Im}[\tilde{\rho}_{21}e^{i\epsilon_c t}] 
& e^{i\epsilon_c t}\left(\tilde{\rho}_{22}-\tilde{\rho}_{11}\right) \\
e^{-i\epsilon_c t}\left(\tilde{\rho}_{11}-\tilde{\rho}_{22}\right) 
& i2\mathrm{Im}[\tilde{\rho}_{12}e^{-i\epsilon_c t}]
\end{pmatrix}\\
&-\frac{\gamma_c}{2}
\begin{pmatrix} -2\tilde{\rho}_{22} & \tilde{\rho}_{12} \\ \tilde{\rho}_{21} & 2\tilde{\rho}_{22} \end{pmatrix}\\
&-\frac{\gamma_z}{2}\bar{n}_z\left(a_za_z^\dagger\tilde{\rho}-2a_z^\dagger\tilde{\rho} a_z+\tilde{\rho} a_za_z^\dagger\right)\\
&-\frac{\gamma_z}{2}\left(\bar{n}_z+1\right)\left(a_z^\dagger a_z\tilde{\rho}-2a_z \tilde{\rho} a_z^\dagger+\tilde{\rho} a_z^\dagger a_z\right).
\end{split}
\label{eq:CyclotronMasterEquation}
\end{equation}
The first line describes time evolution of the density matrix by $\widetilde{V}$. The diagonal terms are 0 because $\ket{1}{n_z}$ and $\ket{2}{n_z}$ are eigenstates of $\widetilde{V}$ for the QND measurement. The non-diagonal terms represents the differing bottle shift for $\ket{1}{n_z}$ and $\ket{2}{n_z}$. The second line describes the electromagnetic cyclotron drive. The third term describes synchrotron radiation from the excited cyclotron state at a rate $\gamma_c$.  The fourth and fifth terms arise from the axial damping and reservoir excitation.  They do not change $P_c$ because they do not change either the cyclotron or spin state. 

The axial damping terms in the master equation (Eq.~(\ref{eq:CyclotronMasterEquation})) generate no coherence between axial states.  Only axially diagonal terms, $\langle i,n_z|\tilde{\rho}|j,n_z\rangle$ are nonzero, where $i$ and $j$ are the labels for the states we consider (Eq.~(\ref{eq:ThreeStates})). The transformation  
\begin{equation}
   p_{ij;n_z}(t)=\langle i,n_z|\tilde{\rho}(t)|j,n_z\rangle \, e^{i(i-j) \epsilon_c t}. 
   \label{eq:pjkn}
\end{equation}  
makes these coefficients carry all the time dependence. Notice that the probability to be in each of the cyclotron and spin states of Eq.~(\ref{eq:Pl}) is also the trace
\begin{eqnarray}
P_l = \sum_{n_z=0}^{\infty} \langle l,n_z | p     \left|l,n_z\right\rangle,
\label{eq:Plp}
\end{eqnarray}
where $p$ has components $p_{jk}$, and $l$ denotes 1, 2, or 3 in Fig.~\ref{fig:quantumstates} and Eq.~(\ref{eq:ThreeStates}).  This is because
the diagonal matrix elements with $i=j$ are equal to the those for the density operator in the Schrodinger picture and the interaction picture.  For the cyclotron excitation being considered in this section, $P_3=0$ because the states $|3,n_z\rangle$ are never populated.  

The differential equations after the transformation are 
\begin{subequations}
\begin{align}
\frac{d}{dt}&p_{11;n_z}(t)\nonumber\\
&=\left[-\gamma_z\left(2\bar{n}_z+1\right)n_z-\gamma_z\bar{n}_z\right]p_{11;n_z}(t)\nonumber\\
&+\gamma_cp_{22;n_z}(t)-\Omega_c\textrm{Im}\left[p_{12;n_z}\right]\nonumber\\
&+\gamma_z\bar{n}_zn_zp_{11;n_z-1}(t)+\gamma_z(\bar{n}_z+1)(n_z+1)p_{11;n_z+1}(t)\label{eq:masterdifferentialA}\\
\frac{d}{dt}&p_{12;n_z}(t)\nonumber\\
&=\big[i\left(-\epsilon_c+\delta_c\left(n_z+\tfrac{1}{2}\right)\right)\nonumber\\
&-\tfrac{1}{2}\gamma_c-\gamma_z(2\bar{n}_z+1)n_z-\gamma_z\bar{n}_z\big]p_{12;n_z}(t)\nonumber\\
&-i\frac{\Omega_c}{2}(p_{22;n_z}(t)-p_{11;n_z}(t))\nonumber\\
&+\gamma_z\bar{n}_zn_zp_{12;n_z-1}(t)+\gamma_z(\bar{n}_z+1)(n_z+1)p_{12;n_z+1}(t)\label{eq:masterdifferentialB}
\\
\frac{d}{dt}&p_{22;n_z}(t)\nonumber\\
&=\left[-\gamma_c-\gamma_z\left(2\bar{n}_z+1\right)n_z-\gamma_z\bar{n}_z\right]p_{22;n_z}(t)\nonumber\\
&+\Omega_c\textrm{Im}\left[p_{12;n_z}\right]\nonumber\\
&+\gamma_z\bar{n}_zn_zp_{22;n_z-1}(t)+\gamma_z(\bar{n}_z+1)(n_z+1)p_{22;n_z+1}(t).
\label{eq:masterdifferentialC}
\end{align}
\end{subequations}
These equations are to be solved for the initial conditions  $p_{12;n_z}(0)=p_{22;n_z}(0)=0$ and $p_{11,n_z}(0) = p_{n_z}(T)$.  Because the states $|3,n_z\rangle$ are never populated, the $p_{j,k} = 0$ when either $j=3$ or $k=3$.

Equations~(\ref{eq:masterdifferentialA}-\ref{eq:masterdifferentialC}) is a matrix equation for the vectors $\vec{p}_{ij}(t)$ with components $p_{ij;n_z}$, 
\begin{subequations}
\begin{align}
\frac{d}{dt}\vec{p}_{11}(t)
&=\mathbf{R}(0,0,0)\, \vec{p}_{11}(t)-\Omega_c\textrm{Im}\left[\vec{p}_{12}(t)\right]+\gamma_c\vec{p}_{22}(t)\\
\frac{d}{dt}\vec{p}_{12}(t)
&=\mathbf{R}(\epsilon_c,\delta_c,\gamma_c)\, \vec{p}_{12}(t)-i\frac{\Omega_c}{2}(\vec{p}_{22}(t)-\vec{p}_{11}(t))\\
\frac{d}{dt}\vec{p}_{22}(t)
&=\mathbf{R}(0,0,2\gamma_c)\, \vec{p}_{22}(t)+\Omega_c\textrm{Im}\left[\vec{p}_{12}(t)\right].
\end{align}
\label{eq:MatrixEquation}%
\end{subequations}
The states $|3,n_z\rangle$ are never populated so all 
The non-zero elements of the  time-independent matrix are 
\begin{subequations}
\begin{align}
\mathbf{R}(\epsilon,\delta,\gamma_c)_{n_z,n_z-1} = & \gamma_z\bar{n}_z n_z\label{eq:Ra}\\
\mathbf{R}(\epsilon,\delta,\gamma_c)_{n_z,n_z} = &   
i\left[-\epsilon+(n_z+\tfrac{1}{2})\delta  \right]-\tfrac{1}{2}\gamma_c\nonumber\\ 
&-\gamma_z(2\bar{n}_z+1)n_z-\gamma_z\bar{n}_z\label{eq:Rb}\\
 \mathbf{R}(\epsilon,\delta,\gamma_c)_{n_z,n_z+1} = & \gamma_z (\bar{n}_z+1) (n_z+1).\label{eq:Rc}
\end{align}
\label{eq:R}%
\end{subequations}
\noindent The initial conditions for the vector differential equations above are  $\vec{p}_{11}(0)=\vec{p}(T)$ and $\vec{p}_{12}(0)=\vec{p}_{22}(0)=0.$

\subsection{Steady-State Cyclotron Line Shape}

After transients have died out in a time 
\begin{equation}
t\gg\gamma_c^{-1},
\end{equation}
a weak drive with $\Omega_c\ll \gamma_c$ produces a steady state in which driven cyclotron excitation balances the incoherent spontaneous emission of synchrotron radiation.  Clearly, 
\begin{eqnarray}
&P_1 = \textrm{Tr}[p_{11}] = &\sum_{n_z} {p}_{11;n_z}(t)\approx \sum_{n_z}p_{n_z}(T) = 1 \\
&P_2 = \textrm{Tr}[p_{22}] = &\sum_{n_z} p_{22;{n_z}}(t) \ll 1
\label{eq:p11}
\end{eqnarray}
and terms involving $\vec{p}_{22}$ are negligibly small compared to those involving $\vec{p}_{11}$.
The resulting steady state, from Eq.~(\ref{eq:MatrixEquation})  with the time derivatives set to zero and the mentioned approximation is described by
\begin{eqnarray}
&\mathbf{R}(\epsilon_c,\delta_c,\gamma_c)\, \vec{p}_{12}+i\frac{\Omega_c}{2}\vec{p}(T)=0\label{eq:SteadyStatep12}
\\
&\mathbf{R}(0,0,2\gamma_c)\, \vec{p}_{22}+\Omega_c\textrm{Im}\left[\vec{p}_{12}\right]=0.
\label{eq:SteadyStatep22}
\end{eqnarray}
The latter can be simplified because 
\begin{equation}
\sum_{{n_z}=0}^{\infty}\left( \mathbf{R}(0,0,2\gamma_c)\, \vec{p}_{22} \right)_{n_z}
=-\gamma_c\textrm{Tr}\left[{p}_{22}\right],
\end{equation}
because axial damping does not change the total population in states $|2,{n_z}\rangle$, and because $\mathbf{R}(0,0,2\gamma_c)$ has a simple structure. 

The result is a steady 
 state probability for weak drive cyclotron excitation, $P_2$ as defined in Eq.~(\ref{eq:Pl}), given by 
 \begin{eqnarray}
 & ~~~~~~~P_2& =P(\Omega_c,\epsilon_c,\delta_c)\\
 &P(\Omega,\epsilon,\delta)\,& \equiv -\frac{\Omega^2}{2\gamma_c}   \textrm{Im}\left[
  \sum_{{n_z}=0}^{\infty}\left(
  i \mathbf{R}(\epsilon,\delta,\gamma_c)^{-1}\vec{p}(T)
  \right)_{n_z}\right].  
\label{eq:P}
\end{eqnarray}
We use arguments without subscripts in $P(\Omega,\epsilon,\delta)$ because this function with other arguments will also describes other steady-state line shapes in what follows.

For the limiting case of a $T=0$ bath, $\bar{n}_z = 0$ and $\vec{p}(T)$  collapses to a single element ${p}_0(T)=1$.  Only the reciprocal of $\mathbf{R}(\epsilon,\delta,\gamma_c)_{0,0} =-i\epsilon +i\delta/2- \frac{1}{2}\gamma_c$ contributes to Eq.~(\ref{eq:P}).  The steady state line shape for a weak drive, $P(\Omega,\epsilon,\delta)$, thus becomes a Lorentzian,  \begin{equation}
P_0(\Omega,\epsilon,\delta) 
= \left( \frac{\Omega}{\gamma_c} \right)^2  
 \frac{ \left(\frac{1}{2}\gamma_c\right)^2   }
{\left(\epsilon-\frac{1}{2}\delta\right)^2 + \left(\frac{1}{2}\gamma_c\right)^2} 
\label{eq:Lorentian}
\end{equation}
in the $T=0$ limit.
The full width at half maximum of this line shape is $\gamma_c$.  The line shape maximum is shifted to $\epsilon=\delta/2$. That this shift is due to the coupling of zero-point fluctuations of the axial motion can be seen by setting $n_z=0$  for the appropriate frequency in Eq.~(\ref{eq:Shifts}).  The steady state probability for being excited with a resonant weak drive is $( \Omega/\gamma_c)^2$.  This is a very small fraction for a weak drive with $\Omega \ll \gamma_c$.  

The symmetric and narrow Lorentzian cyclotron line shape that would pertain for $T=0$ would be ideal experimentally in some respects.  Cavity sideband cooling with a extremely small $\gamma_z$ has been proposed \cite{Review} as way to attain this limit.  This calculation, however, is an investigation of what can be done for a temperature of 0.1 K, an achieved temperature that is close to but not at this limit.  

\subsection{Classical Brownian Motion Line Shape Limit}
\label{sec:ClassicalLimit}

Before the quantum treatment of the coupled spin, cyclotron and axial system presented above, the calculated line shape that was compared to experiment \cite{BrownLineshape,BrownLineshapePRL,Review} assumed the axial detector motion was a classical harmonic oscillation driven by thermal noise. The Brownian motion line shape that resulted from a weak drive is given in terms of a line shape function,
\begin{equation}
\begin{split}
 &\chi (\epsilon,\gamma_z,\bar{n}_z) \\
& = \frac{4}{\pi} \textrm{Re}\Bigg[\frac{\gamma^{\prime}\gamma_z}{\left(\gamma^{\prime}+\gamma_z\right)^2}\sum_{k=0}^{\infty}\frac{\left(\gamma^{\prime}-\gamma_z\right)^{2k}\left(\gamma^{\prime}+\gamma_z\right)^{-2k}}{\left(k+\frac{1}{2}\right)\gamma^\prime+\frac{1}{2}\left(\gamma_c-\gamma_z\right)-i\epsilon}\Bigg],
 \end{split}
\label{eq:classicallineshape1}
\end{equation}
in our notation. (The argument $\epsilon$ for $\chi (\epsilon,\gamma_z,\bar{n}_z)$, like for the function $P$ of Eq.~(\ref{eq:P}), will be equal to one of $\epsilon_c$, $\epsilon_a$ or $\epsilon_s$, as will be specified in context.)   The bath temperature $T$ enters via  
\begin{equation}
 \gamma^\prime=\sqrt{\gamma_z^2+4i\gamma_z\bar{n}_z\delta},
\label{eq:classicallineshape2}
\end{equation}
since this bath temperature determines $\bar{n}_z$, while  $\delta$ and $\epsilon$ are equal to $\delta_c$ ($\delta_a$) and $\epsilon_c$ ($\epsilon_a$) respectively for cyclotron (anomaly) transition. 
The steady state pertains when the transition rate $(\pi/2) \Omega^2 \chi$ (Eq.~(5.19) of \cite{Review}) equals the decay rate $\gamma_c\times P(\Omega,\epsilon,\delta)$.  Thus  
\begin{equation}
P(\Omega,\epsilon,\delta) = \frac{\pi\Omega^2}{2\gamma_c} \chi (\epsilon,\gamma_z,\bar{n}_z)
\end{equation}
is the classical, Brownian motion line shape.

\subsection{Discussion of the Quantum Cyclotron Line Shape}

\begin{figure}[]
    \centering
    \includegraphics[width=\the\columnwidth]{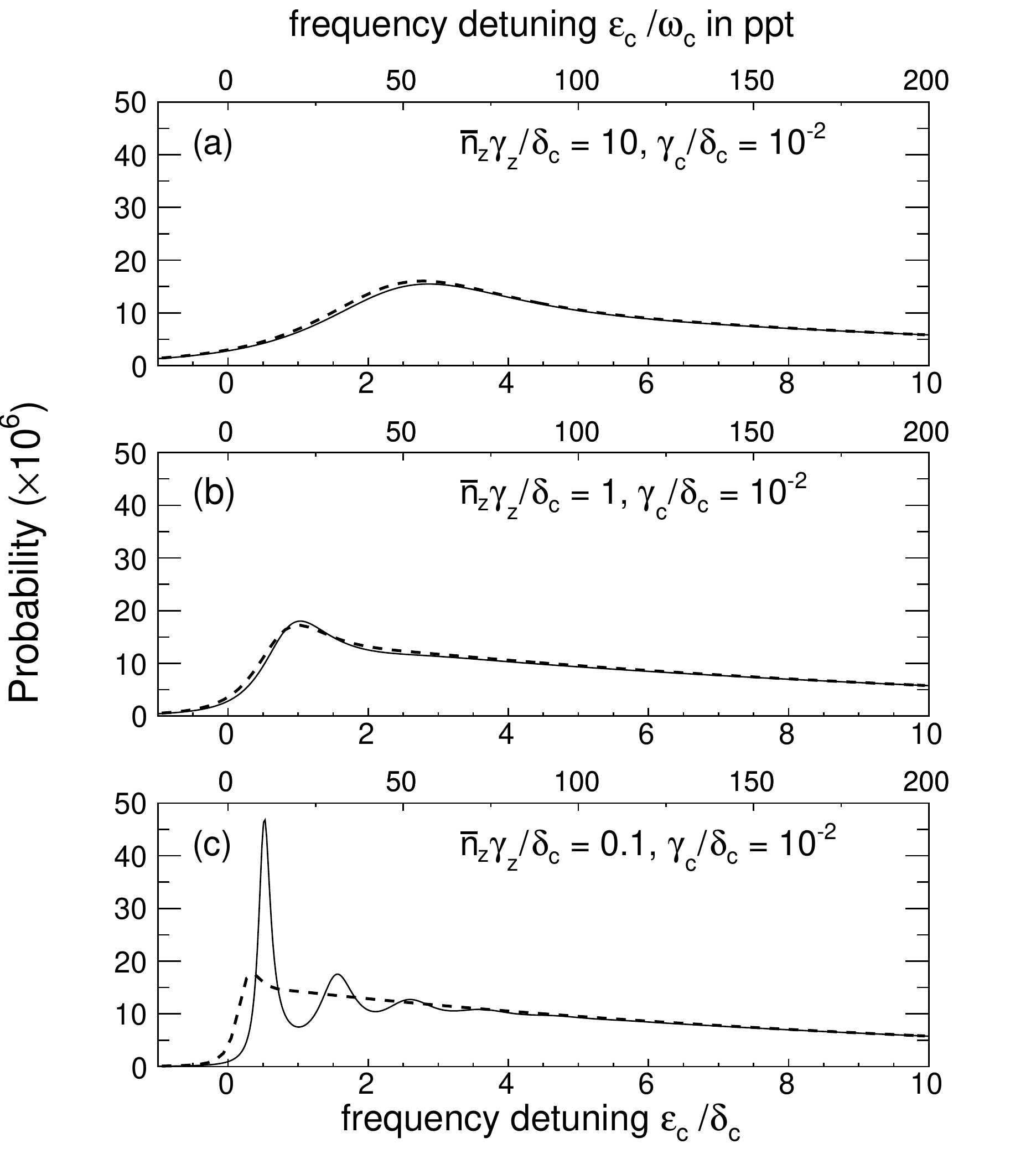}
    \caption{Comparison of quantum calculation (solid) and classical calculation (dashed) with the different $\gamma_z's$  for weak drive ($\Omega_c=0.1\gamma_c$) in cyclotron transition. The two calculations agree when $\bar{n}_z\gamma_z>\delta_c$ as illustrated for damping rates $\gamma_z$ for (a-c) that are 1000, 100, 10 times the value in Table. \ref{table:FrequenciesCompared}.  The  lineshape for the $\gamma_z$ in the table is  presented later in Fig.~\ref{fig:CyclotronLineshape}. }
    \label{fig:ClassicalAndQuantumCyc}
\end{figure}

The quantum steady-state lineshape
 (solid in Fig,~\ref{fig:ClassicalAndQuantumCyc}) is very close to the  Brownian motion steady-state line shape (dashed in Fig,~\ref{fig:ClassicalAndQuantumCyc}) 
 when $\bar{n}_z\gamma_z \gg \delta_c$. This was true for the 2008 measurement
for which $\bar{n}_z\gamma_z\approx6\delta_c$
(using parameters from Table.~\ref{table:FrequenciesComparedOld}).  For weaker axial damping the two line shapes predict every different results, however.

The master equation for driven cyclotron excitation can be solved numerically to reveal the time evolution of the probabilities.  It can also be integrated directly to examine the effect of power broadening when the weak drive condition ($\Omega_c \ll \gamma_c$) is not satisfied. Both will be illustrated.

\begin{figure}[htbp!]
    \centering
    \includegraphics[width=\the\columnwidth]{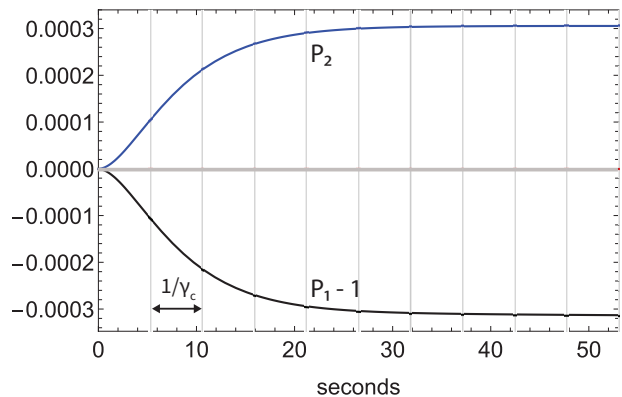}
    \caption{Time evolution in response to a weak and resonant cyclotron drive applied for 10 cyclotron damping times, indicated by vertical gray lines. The probability to be in the $\left|2,n_z\right\rangle$ states (blue) reaches a steady state after transients die out on a time scale give by the cyclotron damping time, $1/\gamma_c$. The  probability to be in the $\left|1,n_z\right\rangle$ states is shown in black with unit probability subtracted out.}
    \label{fig:CyclotronTimeEvolution}
\end{figure}

Figure~\ref{fig:CyclotronTimeEvolution} illustrates the time evolution for a cyclotron drive that is weak ($\Omega_c = 0.1 \gamma_c$, resonant ($\epsilon_c = \delta_c/2$) for the realistic experimental conditions in Table~\ref{table:FrequenciesCompared}. The probability to be in the $\left|2,n_z\right\rangle$ states increases from zero to reach a steady state for $t\gg 1/\gamma_c$.  The cyclotron damping time $1/\gamma_c$ sets the scale for the transients to die out.  The much larger probability to be in the initial $\left|1,n_z\right\rangle$ states stays close to unit probability.  The black curve in the figure shows the small decrease from unit probability needed to conserve probability.   

\begin{figure}[htbp!]
    \centering
    \includegraphics[width=\the\columnwidth]{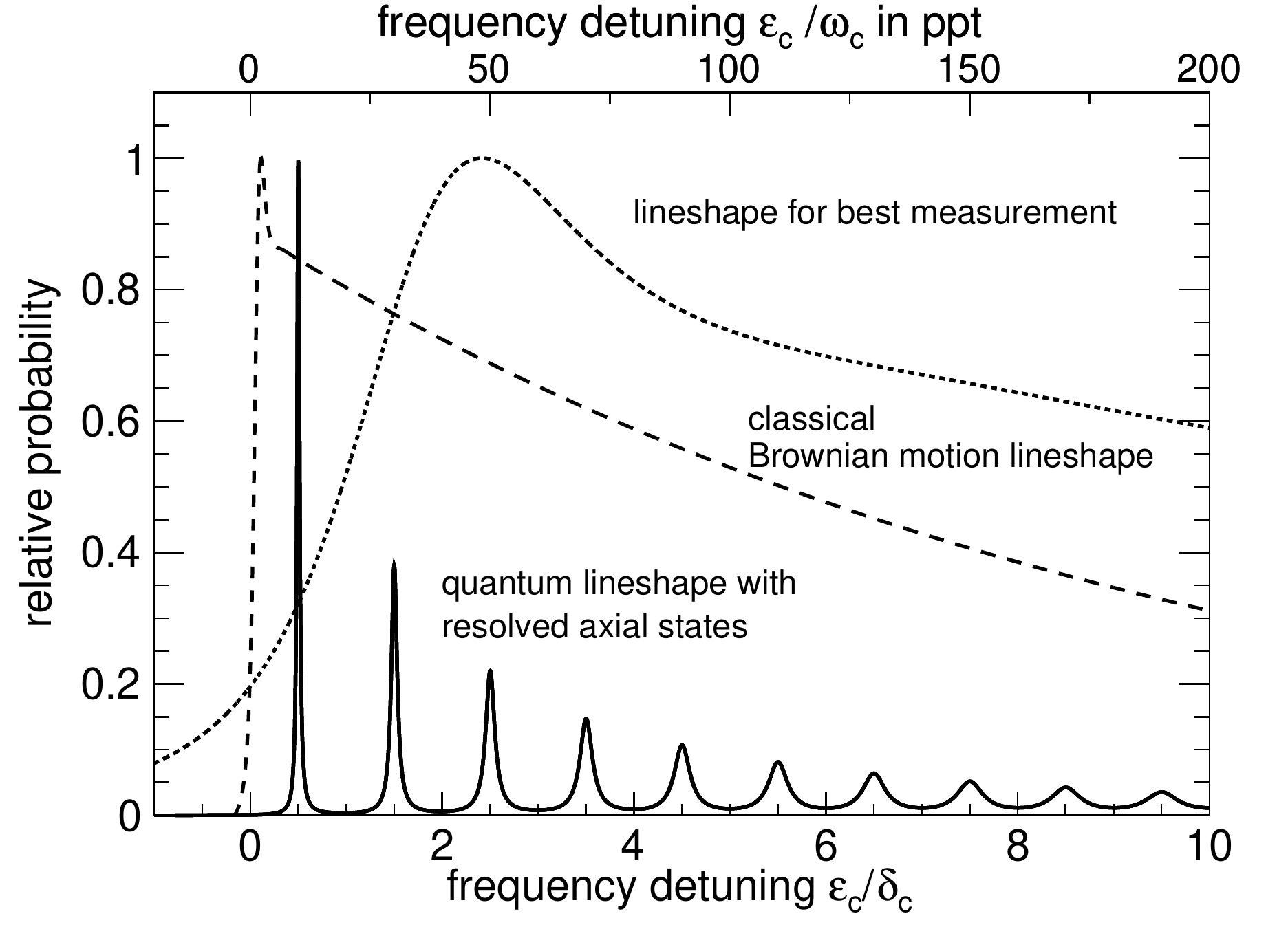}
    \caption{Quantum cyclotron line shape (solid) with clearly resolved axial quantum states (for a weak cyclotron drive with the larges peak normalized to 1) for the quantum calculation (solid), but not for the classical Brownian motion line shape (dashed). The quantum line shape is a huge improvement on the line shape used for the best measurement (dotted).}  
    \label{fig:CyclotronLineshape}
\end{figure}

The resonance line shape for cyclotron excitation is obtained by numerically integrating the master equation from the stated boundary conditions at time $t=0$ to time $t$ for various values of the drive detuning, $\epsilon_c$, as illustrated in Fig.~\ref{fig:CyclotronLineshape}. The probability to be in the states $\left|2,n_z\right\rangle$ at time $t=10\gamma_c^{-1}$ is shown for a cyclotron drive that is weak ($\Omega_c = 0.1 \gamma_c$), for the realistic experimental conditions in Table~\ref{table:FrequenciesCompared}.

\begin{figure}[htbp!]
    \centering
    \includegraphics[width=\the\columnwidth]{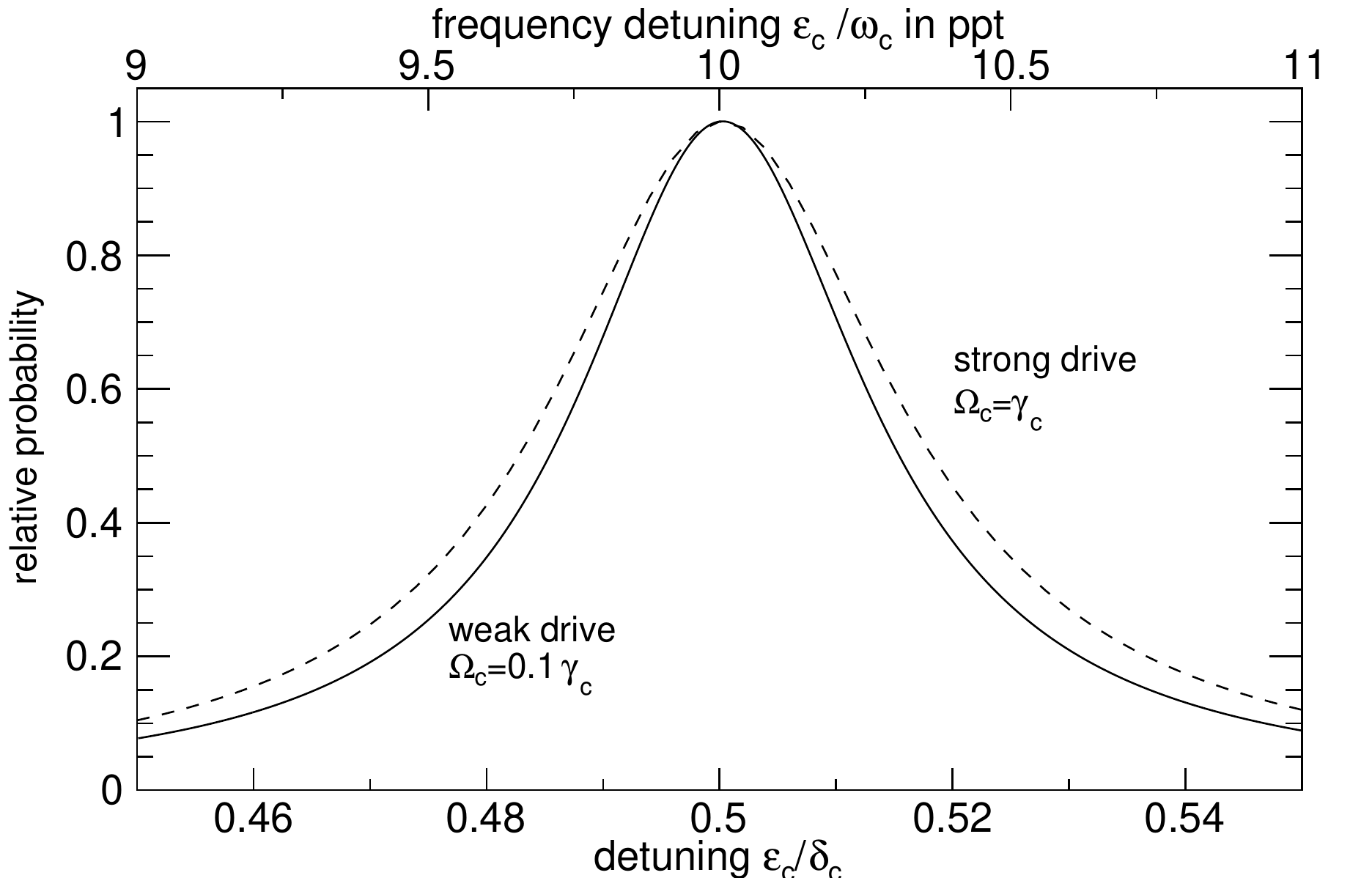}
    \caption{Cyclotron line shape for the resolved $n_z=0$ axial state and a weak drive (solid curve, $\Omega_c=0.1~\gamma_c$) has a full-width at half maximum of about $3 \gamma_c$.  The master equation integrated for 10 cyclotron damping times and the steady-state line shape (solid) coincide. A 10 times stronger drive (dashed, $\Omega_c=\gamma_c)$, only slightly increases the linewidth.}  
    \label{fig:CyclotronFirstPeak}
\end{figure}

The first narrow peak to the left in the figure shows the probability versus drive frequency for making a cyclotron excitation from the cyclotron ground state while the axial motion is in its ground state with $n_z=0$. The series of narrow cyclotron resonances, the first evidence of axial quantization, are for successively higher values of $n_z$ going right.   Resolving these narrow peaks becomes possible only for the small axial damping rate that is now possible experimentally \cite{FanRFSwitch2020}. Each of the peak corresponds to one quantum excitation of cyclotron motion for different $n_z$. This quantum line shape is very different than was observed previously, and it is completely inconsistent with the classical cyclotron line shape, of course. The narrow peaks correspond to resolved quantum states of the axial motion which could not previously be observed.  The left peak is for $n_z=0$, the next for $n_z=1$, and so on. There are many peaks because the average axial quantum number is $\bar{n}_z = 10$ for the experimental conditions in Table ~\ref{table:FrequenciesCompared}.  The individual peaks are resolved because two conditions are met. First, $\bar{n}_z\gamma_z \ll \delta_c$, i.e.\ the width of each axial state, $\bar{n}_z\gamma_z$, is much smaller than the magnetic bottle shift per axial quantum, $\delta_c$. Second, $\gamma_c \ll \delta_c$, i.e.\ the cyclotron damping width is much smaller than the magnetic bottle shift per axial quantum, $\delta_c$. 

The good news from this calculation for potential measurements is how much narrower the $n_z=0$ resonance peak is compared to the cyclotron line shape used for the last electron magnetic moment measurement (dotted in Fig.~\ref{fig:CyclotronLineshape} with experimental parameters in Table~\ref{table:FrequenciesComparedOld}).  In fact, the linewidth of the $n_z=0$ peak is only a factor of 3 larger than the cyclotron linewidth, $\gamma_c$ (Fig \ref{fig:CyclotronFirstPeak}). This is consistent with the indication from Eq.~(\ref{eq:masterdifferentialC}) that the linewidth is of order $\gamma_c + 2\bar{n}_z \gamma_z$. Cavity-inhibition of spontaneous emission makes $\gamma_c$ very small \cite{InhibitionLetter}.  A low temperatures makes $\bar{n}_z$ small, and the previously mentioned new method makes $\gamma_z$ small \cite{FanRFSwitch2020}. 

\begin{table}[htbp!]
\begin{tabular}{c |c |c}
  ang.\ frequency or rate & frequency (Hz) & time constant (s) \\
\hline
\hline
$\delta_a$ & $0.004$ & $40$\\
$\gamma_z$ & $1$& $0.16$\\
$\bar{n}_z\delta_a$ & $0.09$ & $1.7$\\
$\gamma_c$ & $0.03
$ & 6\\
$\bar{n}_z\gamma_z$ & $23$& $0.007$\\
$\delta_c$ & $4$ & $0.04$\\
$\bar{n}_z\delta_c$ & $92$ & $0.0017$
\label{tab:twodrivefrequencyscale}
\end{tabular}
\caption{Hierarchy of angular frequencies and rates used on the best completed experiments \cite{HarvardMagneticMoment2008,HarvardMagneticMoment2011}, to be compared with the previous table.  The axial temperature was also as low as $\bar{n}_z=23$. The numerical values are frequencies in Hz and times in seconds.}
\label{table:FrequenciesComparedOld}
\end{table}

More good news for possible measurements is that the $n_z=0$ peak is quite symmetric about its center frequency.  This is generally a big help in precisely identifying the center frequency of a resonance.  The dotted line in Fig.~\ref{fig:CyclotronLineshape} illustrates the big contrast to the highly asymmetric classical line shape used for previous measurements.  

The small probability, $3.1 \times 10^{-4}$, that a weak cyclotron drive ($\Omega_c=0.1\gamma_c$) will make an excitation within 10 cyclotron damping times (53 seconds) is of some concern.
However, increasing the cyclotron drive strength to $\Omega_c = \gamma_c$ increases the probability for an excitation to $2.2\times 10^{-2}$ while increasing the full linewidth from 3 to only 3.6 cyclotron decay widths (solid and dashed curves in Fig.~\ref{fig:CyclotronFirstPeak}).    This cyclotron linewidth is narrow enough to make possible magnetic moment measurements that are orders of magnitude more accurate than the current limit (assuming the anomaly frequency is determined with a similar accuracy). Because the power broadening is so small, even stronger drives could be used to track a slowly drifting magnetic field \cite{HarvardMagneticMoment2008}.  

The offset of the $n_z=0$ resonance from $\epsilon_c=0$ to $\epsilon_c = \delta_c/2$ is due to the zero point motion of the quantum axial oscillator. Measuring this peak and its neighbor would determine this offset more accurately than is needed for dramatically improved magnetic moment measurements, since these two peaks are spaced by twice the offset. This could be an important new option for  precisely measuring the offset.

In summary, this quantum calculation demonstrates the exciting possibility to fully resolve the axial quantum structure in the cyclotron line shape.  With the achievable reductions in axial damping in Table~\ref{table:FrequenciesCompared}, a cyclotron resonance for a particle in its axial ground state can be fully resolved.  This will make it possible to determine the cyclotron frequency (one of two frequencies needed for a magnetic moment measurement) orders of magnitude more precisely. The broad cyclotron linewidth (larger than $\bar{n}_z \delta_c$) that limited past measurements is essentially removed.

\section{Calculating the Anomaly Line Shape}
\label{sec:AnomalyLineshape}

\subsection{Anomaly Master Equation}

An anomaly drive $V_a$ will transfer population from a thermal distribution of stable, spin-up, cyclotron ground states, $|3,n_z\rangle$ to the unstable states, $|2,n_z\rangle$.  These states will then decay  via the spontaneous emission of synchrotron radiation to the stable spin-down ground states $|1,n_z\rangle$. The attractive feature for measurement is that there is no need to detect an unstable state population before it decays.   

The density operator needed to describe anomaly transitions, \begin{equation}
\tilde{\rho} \equiv \begin{pmatrix} 
\tilde{\rho}_{22} & \tilde{\rho}_{23} \\
\tilde{\rho}_{32} & \tilde{\rho}_{33} 
\end{pmatrix},   
\end{equation}
does not need to include the stable lower states, $|1,n_z\rangle$, though it must include decay to these states.  
It has the upper and lower energy states in the same relative matrix locations as in the previous section. 
What must be calculated is the loss of probability from the initial state during the time that the drive is applied, since this is the probability that a spin-flip transition takes place.  

The master equation in the interaction representation is then a lot like Eq.~(\ref{eq:CyclotronMasterEquation}), with the indices $1\rightarrow 2$ and $2\rightarrow 3$,  
\begin{equation}
\begin{split}
\frac{d}{dt}&  \begin{pmatrix} 
\tilde{\rho}_{22} & \tilde{\rho}_{23} \\
\tilde{\rho}_{32} & \tilde{\rho}_{33} 
\end{pmatrix}\\
=&-{i}\left[a_z^{\dagger}a_z+\frac{1}{2}\right]
\begin{pmatrix} 
0 & -\delta_a\tilde{\rho}_{23} \\
\delta_a\tilde{\rho}_{32} & 0 
\end{pmatrix}\\
&-i\frac{\Omega_a}{2}\begin{pmatrix} 
i2\mathrm{Im}[\tilde{\rho}_{32}e^{i\epsilon_a t}] & e^{i\epsilon_at}\left(\tilde{\rho}_{33}-\tilde{\rho}_{22}\right) \\
e^{-i\epsilon_a t}\left(\tilde{\rho}_{22}-\tilde{\rho}_{33}\right) & i2\mathrm{Im}[\tilde{\rho}_{23}e^{-i\epsilon_a t}]
\end{pmatrix}\\
&-\frac{\gamma_c}{2}\begin{pmatrix} 
2\tilde{\rho}_{22} & \tilde{\rho}_{23} \\
\tilde{\rho}_{32} & 0 
\end{pmatrix}\\
&-\frac{\gamma_z}{2}\bar{n}_z\left(a_za_z^\dagger\tilde{\rho}-2a_z^\dagger\tilde{\rho} a_z+\tilde{\rho} a_za_z^\dagger\right)\\
&-\frac{\gamma_z}{2}\left(\bar{n}_z+1\right)\left(a_z^\dagger a_z\tilde{\rho}-2a_z \tilde{\rho} a_z^\dagger+\tilde{\rho} a_z^\dagger a_z\right).
\end{split}
\label{eq:eqanom}
\end{equation}
The term that is different is the cyclotron damping term that is proportional to $\gamma_c$.  This is because the lower rather than the upper of the two sets of states is unstable. The vanishing element in the matrix comes because the states $|3,n_z\rangle$ do not decay.

The discussion follows essentially the same steps discussed in the previous section. The differential equations  are
\begin{subequations}
\begin{align}
\frac{d}{dt}\vec{p}_{22}(t)
&=\mathbf{R}(0,0,2\gamma_c)\, \vec{p}_{22}(t)-\Omega_a\textrm{Im}\left[\vec{p}_{23}(t)\right]\\
\frac{d}{dt}\vec{p}_{23}(t)
&=\mathbf{R}(\epsilon_a,\delta_a,\gamma_c)\, \vec{p}_{23}(t)-i\frac{\Omega_a}{2}(\vec{p}_{33}(t)-\vec{p}_{22}(t))\\
\frac{d}{dt}\vec{p}_{33}(t)
&=\mathbf{R}(0,0,0)\, \vec{p}_{33}(t)+\Omega_a\textrm{Im}\left[\vec{p}_{23}(t)\right]
,
\end{align}
\label{eq:MatrixEquationAnom}%
\end{subequations}
These equations are to be solved for the initial conditions $\vec{p}_{33}(0)=\vec{p}(T)$ and $\vec{p}_{23}(0)=\vec{p}_{22}(0)=0.$

\subsection{Quasi Steady State Solution}

Coherent, driven anomaly transitions can balance the incoherent spontaneous emission of synchrotron radiation to produce a  quasi steady state.  For a weak drive ($\Omega_a\ll \gamma_c$), the system remains mostly in its initial state, so
\begin{eqnarray}
&P_3 = \textrm{Tr}[p_{33}] = &\sum_{n_z} {p}_{33;n_z}(t)\approx \sum_{n_z}p_{n_z}(T) = 1 \\
&P_2 = \textrm{Tr}[p_{22}] = &\sum_{n_z} p_{22;{n_z}}(t) \ll 1.
\label{eq:LargeRho33}
\end{eqnarray}
The quasi steady state pertains in the time range
\begin{equation}
\gamma_c^{-1}\ll t \ll \gamma_c^{-1} \left(\frac{\gamma_c}{\Omega_a}\right)^2.\label{eq:TimeRange}
\end{equation}
The time must be long enough for transients to die out. It must be short enough that Eq.~(\ref{eq:LargeRho33}) remains valid, with the upper time limit justified presently.  What is detected is the probability $P_1$ to end up in the spin-down cyclotron ground state. This probability increases as 
\begin{equation}
    \frac{dP_1}{dt}=\gamma_cP_2
    \label{eq:dP1dt}
\end{equation}
via synchrotron emission from $P_2$ at rate $\gamma_c$.

For the quasi steady state, the time derivatives of $\vec{p}_{22}$ and $\vec{p}_{23}$ are set to zero in Eq.~(\ref{eq:LargeRho33}), though that of $\vec{p}_{33}$ is not, so that   
\begin{eqnarray}
&\mathbf{R}(\epsilon_a,\delta_a,\gamma_c)\, \vec{p}_{23}-i\frac{\Omega_a}{2}\vec{p}(T)=0\label{eq:SteadyStatep12}
\\
&\mathbf{R}(0,0,2\gamma_c)\, \vec{p}_{22}-\Omega_a\textrm{Im}\left[\vec{p}_{23}\right]=0.
\label{eq:SteadyStatep22}
\end{eqnarray}
Because $\mathbf{R}(0,0,2\gamma_c)$ has a simple structure,
\begin{equation}
P_2=\textrm{Tr}\left[{p}_{22}\right]
= -\frac{1}{\gamma_c}
\sum_{{n_z}=0}^{\infty}\left(\mathbf{R}(0,0,2\gamma_c)\, \vec{p}_{22}\right)_{n_z}.
\label{eq:simplified22}
\end{equation} 
Eqs.~(\ref{eq:dP1dt}-\ref{eq:simplified22}) together give a quasi steady state rate
\begin{equation}
    \frac{dP_1}{dt}= \gamma_cP(\Omega_a,\epsilon_a,\delta_a)
    \label{eq:dP1dtSteady}
\end{equation}
that is the same function that described the steady state for cyclotron excitation Eq.~(\ref{eq:P}) multiplied by $\gamma_c$.  With anomaly arguments rather than cyclotron arguments, however, the function takes an entirely different shape.  

When the drive is applied for time $t_d$ and then turned off, the probability $P_1$ eventually becomes the integral of $dP_1/dt$ at time $t_d$ plus $P_2(t_d)$, because the latter probability is transferred to the spin-down ground state by spontaneous emission from the cyclotron excited state. 
Approximating $dP_1/dt$ with the quasi steady state value in Eq.~(\ref{eq:dP1dtSteady}) gives 
\begin{equation}
 P_{1;\mathrm{total}} \approx \left(t_d\gamma_c+1\right)P(
 \Omega_a,\epsilon_a,\delta_a).  
\end{equation}
This slightly overstates the transition probability because $dP_1/dt$ increases before the quasi steady state is established, but the line shape is approximately right.

The $T=0$ limit of the quasi-steady-state anomaly line shape for a weak drive becomes a Lorentzian, $\left(t_d\gamma_c+1\right)P_0(\Omega_a,\epsilon_a,\delta_a)$, similar to what was discussed for cyclotron resonance.  On resonance, the quasi steady state probability to be in state $\ket{2}{0}$ at $T=0$ is  $( \Omega_a/\gamma_c)^2$. This is extremely small for a weak anomaly drive with $\Omega_a \ll \gamma_c$.  For the cases we consider, with temperatures not far from 0, we expect that the rate to transfer population from the initial $\ket{3}{n_z}$ states to the final $\ket{1}{n_z}$ states goes as this small probability times the rate $\gamma_c$ to decay form $\ket{2}{n_z}$ to  $\ket{1}{n_z}$. The population transfer will be small (as needed to have a quasi steady state) as long as the time is short compared to the inverse of this rate, which gives the upper time limit in Eq.~(\ref{eq:TimeRange}).

\subsection{Discussion of the Anomaly Line Shape}

\begin{figure}[htbp!]
      \includegraphics[width=\the\columnwidth]{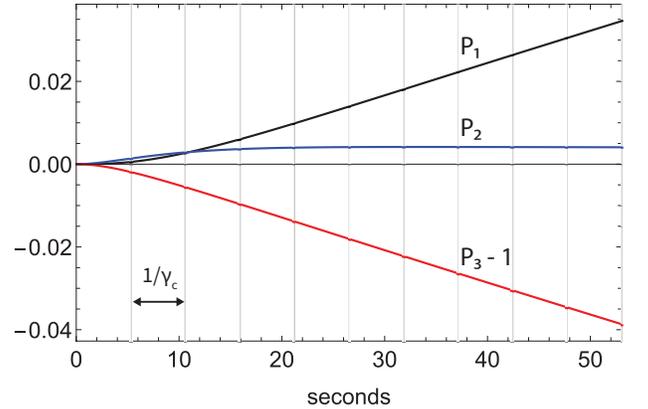} \hfill
    \caption{Time evolution in response to a weak ($\Omega_a = 0.1 \gamma_c$) and resonant anomaly drive (at a drive detuning $\epsilon_a = 5 \delta_a$) with a cyclotron damping times indicated by vertical grid lines.}    \label{fig:AnomalyTimeEvolution}
\end{figure}

Figure~\ref{fig:AnomalyTimeEvolution} is a numerical solution to the master equation for an anomaly drive that is weak ($\Omega_a = 0.1 \gamma_c$) and resonant (at a drive detuning $\epsilon_a = 5 \delta_a$) for the realistic experimental conditions in Table~\ref{table:FrequenciesCompared}. 
The probability $P_2$ increases from zero to reach a quasi steady state in several cyclotron damping times, whereupon the probability $P_1$ increases linearly.  The probability to be in the initial spin-up ground state, $P_3$, decreases only slightly from unity to conserve probability.  

\begin{figure}[htbp!]
    \centering
    \includegraphics[width=\the\columnwidth]{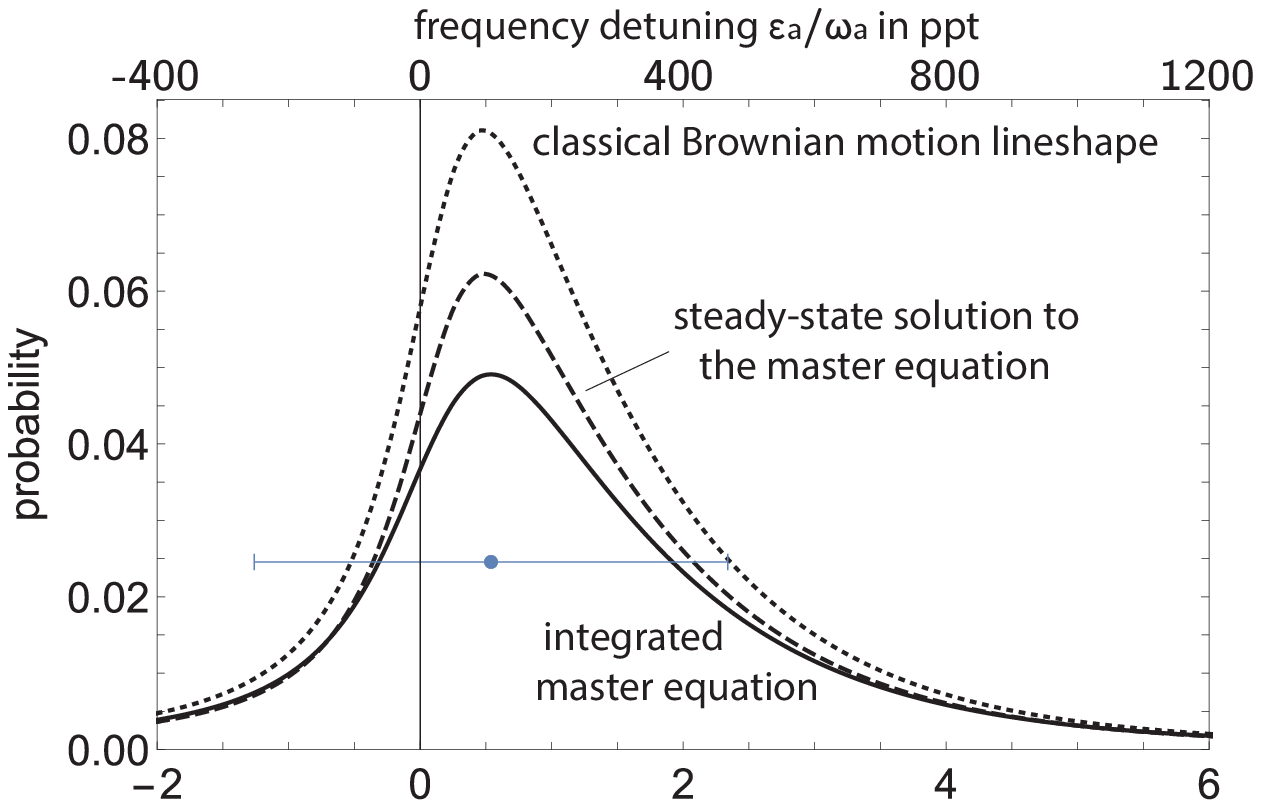}
     \includegraphics[width=\the\columnwidth]{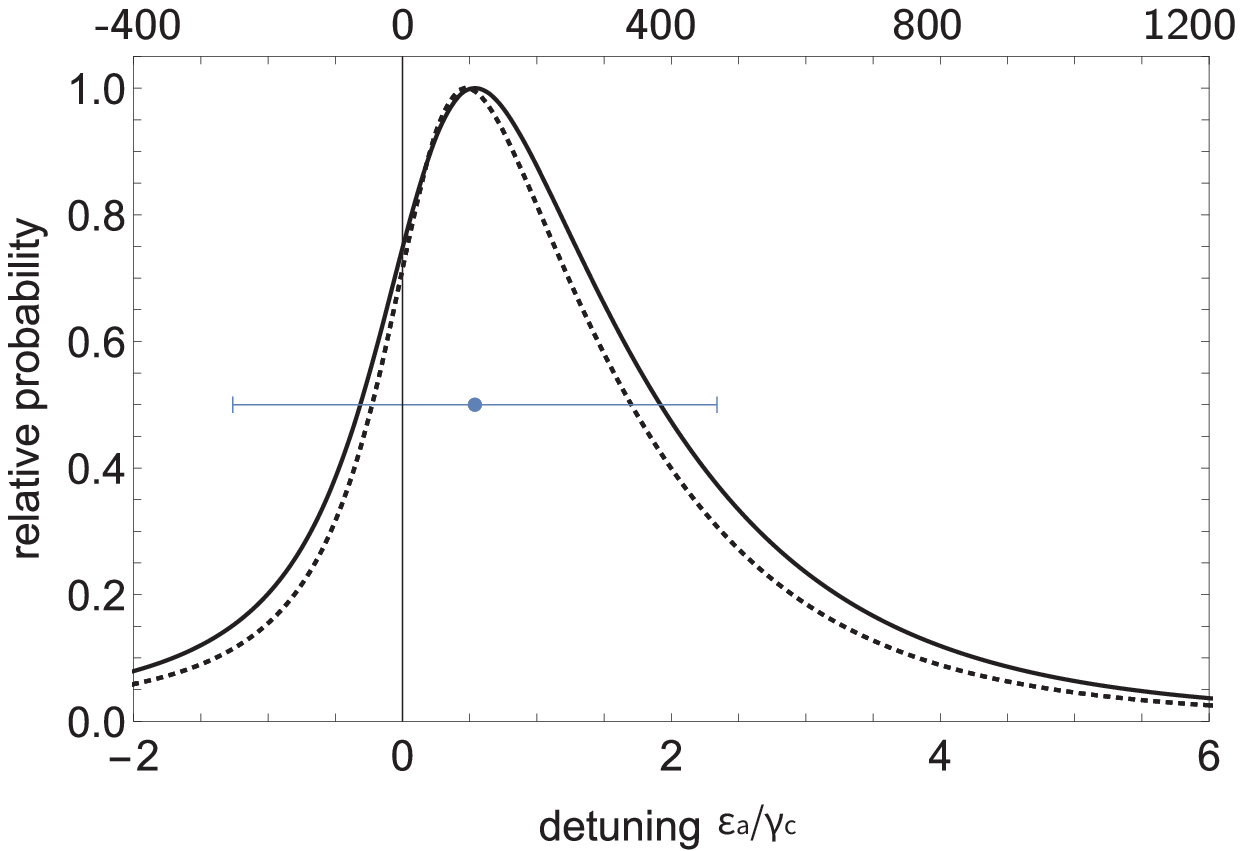}
    \caption{(a) Anomaly line shape for spin flip transition induced by a weak anomaly drive.  
    The integrated solution of the master equation for time $10/\gamma_c$ (solid) is compared to the quasi steady state solution (dashed) and the classical Brownian motion line shape (dotted).   (b) The integrated and steady-state solutions coincide when normalized to their peak probability.  These line shape is much narrower than the $\pm 300$ ppt uncertainty of the best measurement (represented by the "error bar").   
    }    
    \label{fig:AnomalyLineshape}
\end{figure}

The resonance line shapes for driven anomaly transitions in Fig.~\ref{fig:AnomalyLineshape} are for a weak drive ($\Omega_a = 0.1 \gamma_c$) and  the realistic experimental conditions in Table~\ref{table:FrequenciesCompared}.  
The probability $P_{1;\mathrm{total}} = P_1(t_d) + P_2(t_d)$ is plotted versus the detuning  $\epsilon_a$ of the drive from $\omega_a$. The sold curve is obtained by numerically integrating the master equation for ten cyclotron damping times,  $t=10\gamma_c^{-1}$.  The quasi steady state solution (dashed) overestimates the probability because it takes some time to increase the transition rate to the steady state.  However, the normalized line shapes in Fig.~\ref{fig:AnomalyLineshape}b shows that the quasi steady state line shape correctly predicts the shape.  

\begin{figure}[]
    \centering
    \includegraphics[width=\the\columnwidth]{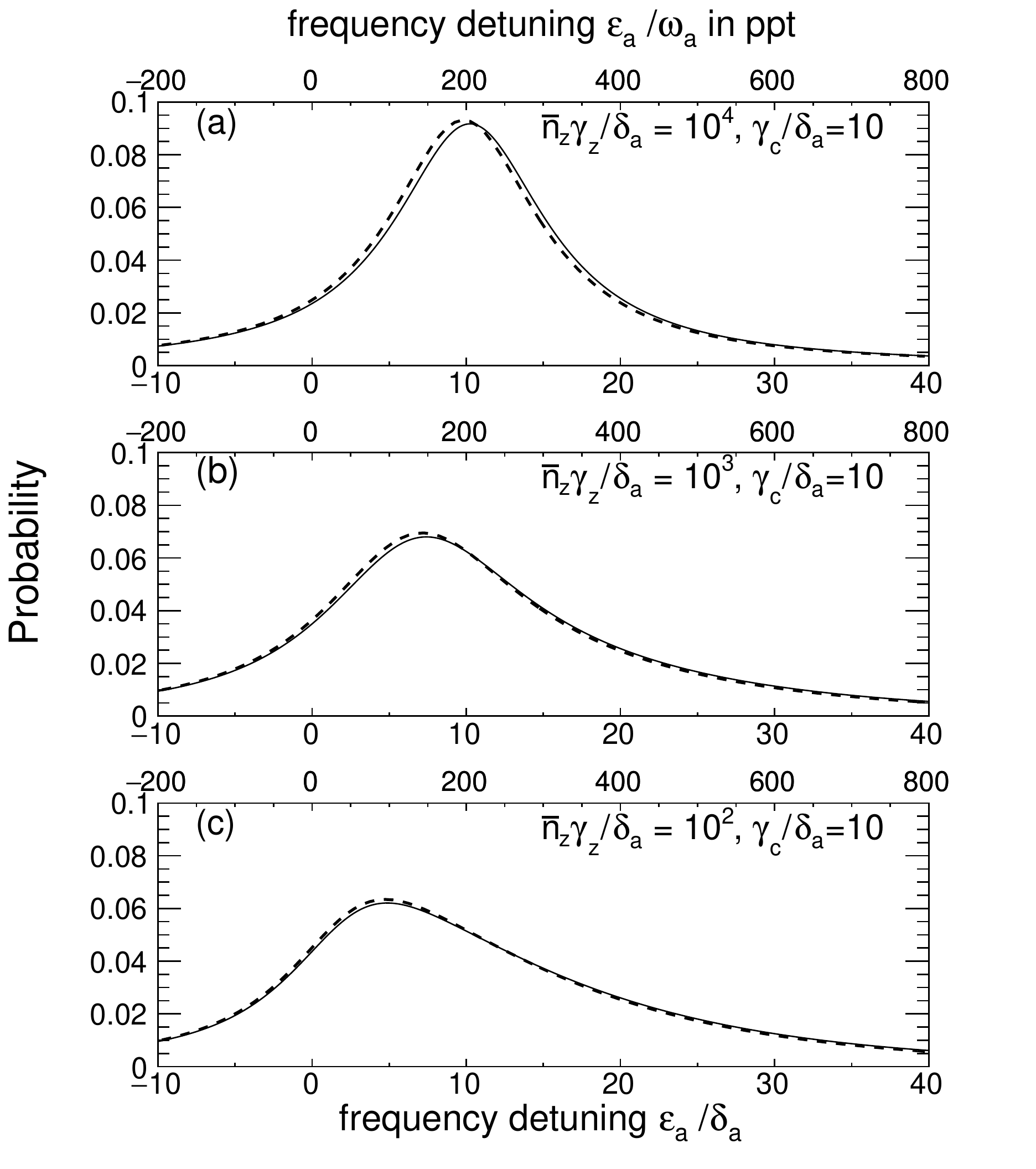}
    \caption{Comparison of quantum calculation (solid) and classical calculation (dashed) with the different $\gamma_z's$  for weak drive ($\Omega_c=0.1\gamma_a$) and 10$\gamma_c^{-1}$ drive time in anomaly transition. The damping rates $\gamma_z$ for (a-c) are 1000, 100, 10 times the value in Table. \ref{table:FrequenciesCompared}, respectively.}
    \label{fig:ClassicalAndQuantumAnom}
\end{figure}

In Fig.~\ref{fig:AnomalyLineshape}, the classical Brownian motion line shape (dotted) is remarkably close to the solution to the master equation obtained by direct integration (solid), quite unlike the case for the cyclotron line shape.  Figure \ref{fig:ClassicalAndQuantumAnom}  compares quantum and classical calculations with three realizable values of $\gamma_z$. For the best measurement \cite{HarvardMagneticMoment2008}, with  $\bar{n}_z\gamma_z=6\times10^3\delta_a$ (Table.~\ref{table:FrequenciesComparedOld}), the two calculations predicts same line shape.

What is so different from the case of the cyclotron lineshape is that the circumvention of detection backaction that was possible in the cyclotron case is not possible for the anomaly lineshape.  The axial quantum states are not resolved within the anomaly lineshape for the realistic parameters of Table \ref{table:FrequenciesCompared}.  The reason is that the anomaly frequency shift per axial quantum of excitation is about 10 times smaller than both the cyclotron damping width $\gamma_c$ and the axial decoherence width $\bar{n}_z\gamma_z$.  The anomaly frequency must be extracted from a resonance line with a calculated linewidth that is about $2.2~\gamma_c$.  The shape is slightly asymmetric with a tail toward higher frequencies because more populated axial states have $n_z > \bar{n}_z$. 

The good news that the calculation nonetheless brings for measurements is that the predicted linewidth (for the realistic conditions of Table ~\ref{table:FrequenciesCompared}) is much narrower than previously realized. The ``error bar'' in the figure corresponds to the $\pm 300$ ppt uncertainty (ppt = 1 part in $10^{12}$) of the most accurate measurement to date \cite{HarvardMagneticMoment2008,HarvardMagneticMoment2011}.  
The full halfwidth of the predicted lineshape is 60\% of the error bar, so a modest  linesplitting of only a factor of 6 would suffice for a ten times more accurate measurement of the electron magnetic moment.

\subsection{Temperature and Damping Dependence}

Once the detection backaction is circumvented \cite{Fan2020BackActionPRL}, determining $\omega_c$ from the cyclotron lineshape should no longer be the leading impediment to measuring the electron and positron magnetic moments orders of magnitude more precisely than has been possible. Since a similar method is not available for measuring the anomaly frequency $\omega_a$, this promises to be the central challenge for future measurements.  The lineshape prediction discussed in the previous section suggests the possibility for a ten-fold improvement if the experimental parameters that currently seem feasible (Tab. \ref{table:FrequenciesCompared}) are realized.  The purpose of this section is to search for possible reductions in anomaly linewidth that may be possible with reductions in axial temperature,  cyclotron damping rate, and axial damping rate beyond the values in the table.  

\begin{figure}[htbp!]
    \centering
    \includegraphics[width=\the\columnwidth]{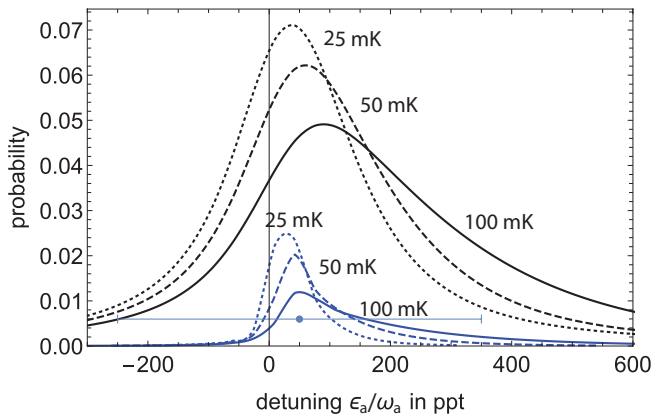}
    \caption{Probability of a spin-flips caused by a driven anomaly transitions for cyclotron and axial reservoir temperatures of 100 mK (solid), 50 mK (dashed) and 25 mK (dotted).  The black curves are the experimentally accessible parameters in Table \ref{table:FrequenciesCompared}.  The blue curves are for a ten-fold reduction in the cyclotron damping rate below the value in the table. 
    }    
    \label{fig:FutureAnomalyPossibilities}
\end{figure}

Fig.~\ref{fig:FutureAnomalyPossibilities} shows anomaly line shapes for a weak drive ($\Omega_a=\gamma_c/10$) for temperatures of 100 mK (black solid), 50 mK (black dashed) and 25 mK (black dotted). The other parameters used are from Table \ref{table:FrequenciesCompared}).  The most accurate measurement was done at an ambient temperature of 100 mK  \cite{HarvardMagneticMoment2008}, with a demonstrated electron cyclotron temperature of 100 mK and a demonstrated axial temperature as low as 230 mK.  The temperature in the table assumes that with better detectors under development, that the latter temperature can be reduced to the ambient.  However, 
dilution refrigerators can reach lower temperatures if the heat load can be made low enough. Also, cavity-sideband cooling is a possible method to reduce the axial temperature below the ambient apparatus temperature \cite{atomsNewMeasurement2019}. The anomaly lineshapes clearly reduce and the lines become more symmetric for lower axial temperatures.  

The blue curves in Fig.~\ref{fig:FutureAnomalyPossibilities} show the large anomaly lineshape reduction that comes from lowering the cyclotron radiation rate by a factor of ten. The most accurate experiment achieved the low damping rate in the table by using a microwave cavity to suppress the spontaneous emission of synchrotron radiation \cite{InhibitionLetter} by a factor of about 200.  An lower loss microwave cavity could further reduce the cyclotron damping rate to produce the narrower lineshapes.   This would slow the measurement because it takes several cyclotron damping times for the population excited to states $\ket{2}{n_z}$ to decay to the ground state, but the damping rate could varied by tuning $\omega_c$ closer or further from cavity microwave resonances \cite{HarvardMagneticMoment2011}.  

Reducing the axial temperature without reducing $\gamma_c$ reduces the linewidth somewhat.  A bigger consequence is that the doing so reduces the asymmetry of the line shape, which should make it possible to identify the resonance frequency more reliably.  The effects of the axial damping rate have also been investigated. Further reductions in the axial damping rate do not noticeably change any of the curves in Fig. \ref{fig:FutureAnomalyPossibilities}.      

\begin{figure}[htbp!]
    \centering
    \includegraphics[width=\the\columnwidth]{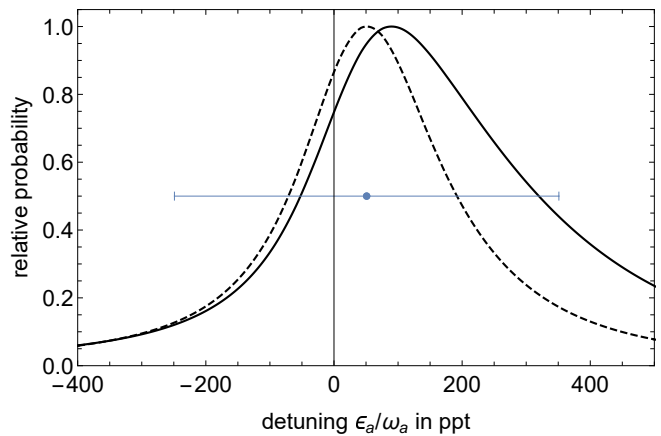}
        \caption{Probability of a spin-flips caused by a driven anomaly transitions for a 100 mK temperature bath (solid). If the axial motion is initially cooled to the $T=0$ limit so that only $n_z=0$ is initially populated, then the linewidth narrows from $\pm 190$ ppt to $\pm 130$ ppt, becomes more symmetric and has a slightly smaller offset frequency.  
    }    
    \label{fig:AnomalyCavitySidebandCooling}
\end{figure}

The possibly to use cavity sideband cooling of the axial motion has been mentioned as a possible route to narrower resonance linewidths \cite{atomsNewMeasurement2019}.  Once the cooling is stopped, the axial motion would then reequilibrate at the bath temperature at a rate $\gamma_z$.  This is not a steady state, of course, but we can investigate the possibility by directly integrating the master equation.
Fig.~\ref{fig:AnomalyCavitySidebandCooling} shows the probability of a spin-flip caused by a weak anomaly drive ($\Omega_c=0.1 \gamma_c$) applied for a 100 mK temperature bath (solid).  For this illustration, the axial motion is initially assumed to be cooled to the $T=0$ limit so that only the lowest axial quantum state is initially populated.  This causes the linewidth to narrow from $\pm 190$ ppt to $\pm 130$ ppt (dashed). The line shape also is more symmetric about its center, and the offset frequency is smaller. 
The drive is applied for time $10/\gamma_c$ in this illustration, which is one axial damping time $1/\gamma_z$.  For the parameters we are using for this illustration (Table~\ref{table:FrequenciesCompared}), the linewidth gets broader for shorter driving times because of the limited drive duration, so narrower resonances would come for a smaller $\gamma_z$.  

\begin{figure}[htbp!]
    \centering
    \includegraphics[width=\the\columnwidth]{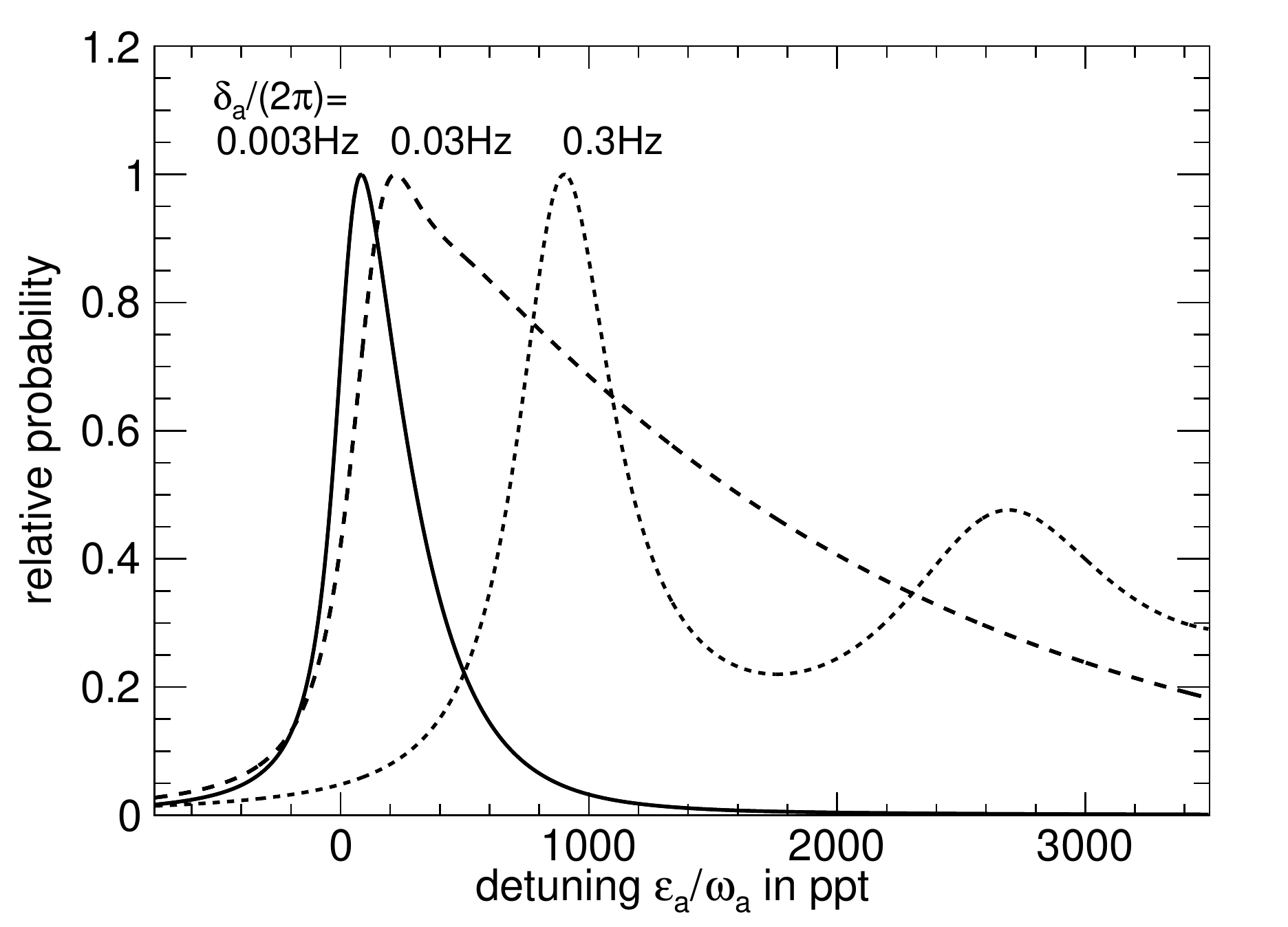}
        \caption{Probability of a spin-flips caused by a driven anomaly transitions for the typical magnetic bottle in Table \ref{table:FrequenciesCompared} (solid), 10 times larger bottle (dashed) and 100 times larger bottle(dotted). 
    }    
    \label{fig:Anomaly:LargeBottle}
\end{figure}

Achieving detection circumvention by resolving the axial states in the anomaly line shape, just as for the cyclotron line shape, would require increasing the bottle shift $\delta_a$ per axial quantum by a factor of 100 or more. This is to make the bottle shift much larger than both the axial decoherence width ($\bar{n}_z\gamma_z$) and the cyclotron damping width ($\gamma_c$).  The solid curve in Fig. \ref{fig:Anomaly:LargeBottle} shows the anomaly line shape for the parameters in Table \ref{table:FrequenciesCompared}. The anomaly lineshape broadens for a 10 times larger bottle gradient.  For a 100 times larger bottle the line begins to separate into peaks that correspond to individual axial quantum states.  Magnetic bottle gradients of the size needed have been produced, but only for Penning traps that are smaller than is otherwise desirable for electron and positron measurements \cite{PbarMagneticMoment2013,ProtonMagneticMoment,OneProtonSpinFlipHarvard,MainzSpinFlips,BASEQOverM2015,BasePbarMagneticMoment,BaseProtonMagneticMomemt}.  However, the figure illustrates that resolving the axial quantum states is not an advantage in that the linewidth of the lowest resolved peak is a bit bigger than the anomaly linewidth already considered. As mentioned above, the linewidth from both the cyclotron damping and the axial decoherence broadening do not decrease with  bottle gradient size.

This cursory survey of anomaly line shapes reveals  no obvious way to make a single large additional reduction in the anomaly linewidth beyond the order of magnitude that has been discussed.

\section{Directly Driven Spin Flips}
\label{sec:DirectSpinFlips}
A spin drive $V_s$ transfers population between the spin-down and spin up cyclotron ground states, $|1,n_z\rangle$ and $|3,n_z\rangle$, both of which are stable. In this section, we apply a spin-flip drive with a Rabi frequency $\Omega_s$ to an initial population in $|1\rangle$, with no cyclotron or anomaly drives (i.e.\  $\Omega_c=0$ and $\Omega_a=0$).  If only one axial detection state was populated this would be the prototypical ``Rabi flopping'' of the two states of a spin qubit.  A distribution of axial detection states has a backaction that makes a superposition of spin frequencies, the effect of which is calculated and discussed here.   

The master equation for the density operator describing this case, in the interaction representation, is 
\begin{equation}
\begin{split}
\frac{d}{dt} &\begin{pmatrix} \tilde{\rho}_{11} & \tilde{\rho}_{13} \\ \tilde{\rho}_{31} & \tilde{\rho}_{33} \end{pmatrix} = -{i}\left(a_z^{\dagger}a_z+\tfrac{1}{2}\right)
\begin{pmatrix} 0 & -\delta_s\tilde{\rho}_{13} \\ \delta_s\tilde{\rho}_{31} & 0 \end{pmatrix}\\
&-i\frac{\Omega_s}{2}
\begin{pmatrix}  
i2\mathrm{Im}[\tilde{\rho}_{31}e^{i\epsilon_s t}] 
& e^{i\epsilon_s t}\left(\tilde{\rho}_{33}-\tilde{\rho}_{11}\right) \\
e^{-i\epsilon_s t}\left(\tilde{\rho}_{11}-\tilde{\rho}_{33}\right) 
& i2\mathrm{Im}[\tilde{\rho}_{13}e^{-i\epsilon_s t}]
\end{pmatrix}\\
&-\frac{\gamma_z}{2}\bar{n}_z\left(a_za_z^\dagger\tilde{\rho}-2a_z^\dagger\tilde{\rho} a_z+\tilde{\rho} a_za_z^\dagger\right)\\
&-\frac{\gamma_z}{2}\left(\bar{n}_z+1\right)\left(a_z^\dagger a_z\tilde{\rho}-2a_z \tilde{\rho} a_z^\dagger+\tilde{\rho} a_z^\dagger a_z\right).
\end{split}
\label{eq:SpinMasterEquationInitial}
\end{equation}
This master equation is the same as for driven cyclotron transitions (Eq.~(\ref{eq:CyclotronMasterEquation})) except that the state $\vert2\rangle$ is replaced by  $\vert3\rangle$ and the damping term $\gamma_c$ is replaced by $\gamma_s\approx0$ (see Tab.~\ref{table:Frequencies}).

The master equation can be solved exactly in the same way as the cyclotron transition. We assume the initial population is distributed in the state $\vert1,n_z\rangle$ with the Boltzmann distribution as Eq.~(\ref{eq:11distribution}).  In vector form, 
Eq.~(\ref{eq:SpinMasterEquationInitial}) is 
\begin{subequations}
\begin{align}
\frac{d}{dt}\vec{p}_{11}(t)
&=\mathbf{R}(0,0,0)\, \vec{p}_{11}(t)-\Omega_s\textrm{Im}\left[\vec{p}_{13}(t)\right]\label{eq:MatrixEquationSpinA}
\\
\frac{d}{dt}\vec{p}_{13}(t)
&=\mathbf{R}(\epsilon_s,\delta_s,0)\, \vec{p}_{13}(t)-i\frac{\Omega_s}{2}(\vec{p}_{33}(t)-\vec{p}_{11}(t))\label{eq:MatrixEquationSpinB}
\\
\frac{d}{dt}\vec{p}_{33}(t)
&=\mathbf{R}(0,0,0)\, \vec{p}_{33}(t)+\Omega_s\textrm{Im}\left[\vec{p}_{13}(t)\right].
\label{eq:MatrixEquationSpinC}
\end{align}
\label{eq:MatrixEquationSpin}%
\end{subequations}
In general, these equations will be solved for an initial values of the density operators at $t=0$.

Since the damping between the spin states $\vert1\rangle$ and $\vert3\rangle$ is essentially zero, the steady state is not as obvious as in the case for cyclotron transitions. The axial decoherence term $\bar{n}_z\gamma_z$ in Eq.~(\ref{eq:MatrixEquationSpin}) does not induce transition between $\vert1\rangle$ and $\vert3\rangle$, but there is still a useful quasi steady state solutions for the ``weak" drive limit, $\Omega_s \ll \bar{n}_z\gamma_z$.

\subsection{Steady State}

If the spin-flip drive is applied for a long time, $t\gg \bar{n}_z \gamma_z/\Omega_s^2,$ and $\bar{n}_z\gamma_z \ne 0$, there is a steady state described by setting the time derivatives in Eqs.~(\ref{eq:MatrixEquationSpin}) to zero, 
\begin{subequations}
\begin{align}
&\mathbf{R}(0,0,0)\, \vec{p}_{11}(t)-\Omega_s\textrm{Im}\left[\vec{p}_{13}(t)\right]=0\label{eq:MatrixEquationSpinSteadyA}
\\
&\mathbf{R}(\epsilon_s,\delta_s,0)\, \vec{p}_{13}(t)-i\frac{\Omega_s}{2}(\vec{p}_{33}(t)-\vec{p}_{11}(t))=0\label{eq:MatrixEquationSpinSteadyB}
\\
&\mathbf{R}(0,0,0)\, \vec{p}_{33}(t)+\Omega_s\textrm{Im}\left[\vec{p}_{13}(t)\right]=0.
\label{eq:MatrixEquationSpinSteadyC}
\end{align}
\label{eq:MatrixEquationSpinSteady}%
\end{subequations}
Summing  Eqs.~(\ref{eq:MatrixEquationSpinSteadyA})  and (\ref{eq:MatrixEquationSpinSteadyC}) over $n_z$, and using
\begin{equation}
\sum_{{n_z}=0}^{\infty}\left( \mathbf{R}(0,0,0)\, \vec{p}_{11} \right)_{n_z}
=\sum_{{n_z}=0}^{\infty}\left( \mathbf{R}(0,0,0)\, \vec{p}_{33} \right)_{n_z}
=0,
\end{equation}
gives $\vec{p}_{13}=0$  and equal populations
\begin{equation}
\sum_{{n_z}=0}^{\infty}\left( \vec{p}_{11} \right)_{n_z}
=\sum_{{n_z}=0}^{\infty}\left( \vec{p}_{33} \right)_{n_z}
=\frac{1}{2}.
\label{eq:SteadyStatePopulationSpin}
\end{equation}
of spin up and spin down states. The interaction of the axial motion with its thermal reservoir produces a spread of Rabi Flopping frequencies, averages out the net Rabi flopping between the two spin states.

\subsection{Quasi Steady State}

A steady state with equal spin up and spin down populations is not useful for determining the spin frequency $\omega_s$.  There is a quasi steady state, however.  
For a ``weak drive" with $\Omega_s\ll\bar{n}_z\gamma_z$ that does not appreciably change the initial Boltzmann distribution $\vec{p}_{11} \approx \vec{p}(T)$ of axial states, there is an excitation rate
\begin{equation}
 \frac{dP_3}{dt} =\frac{d}{dt}\left[\sum_{{n_z}=0}^{\infty}\vec{p}_{33;n_z}(t)\right].
\end{equation}
The drive must be applied for a time in the range 
\begin{equation}
    \left(\bar{n}_z\gamma_z\right)^{-1}\ll t \ll\left(\frac{\Omega_s^2}{\bar{n}_z\gamma_z}\right)^{-1},
\end{equation}
long compared to the dephasing time $\bar{n}_z \gamma_z$ but short compared to the time for approaching the steady-state with equal spin-up and spin-down populations.  Fig.~\ref{Fig:SpinFlipLineShape} shows an example of the time evolution.  

\begin{figure}
    \centering
    \includegraphics[width=\the\columnwidth]{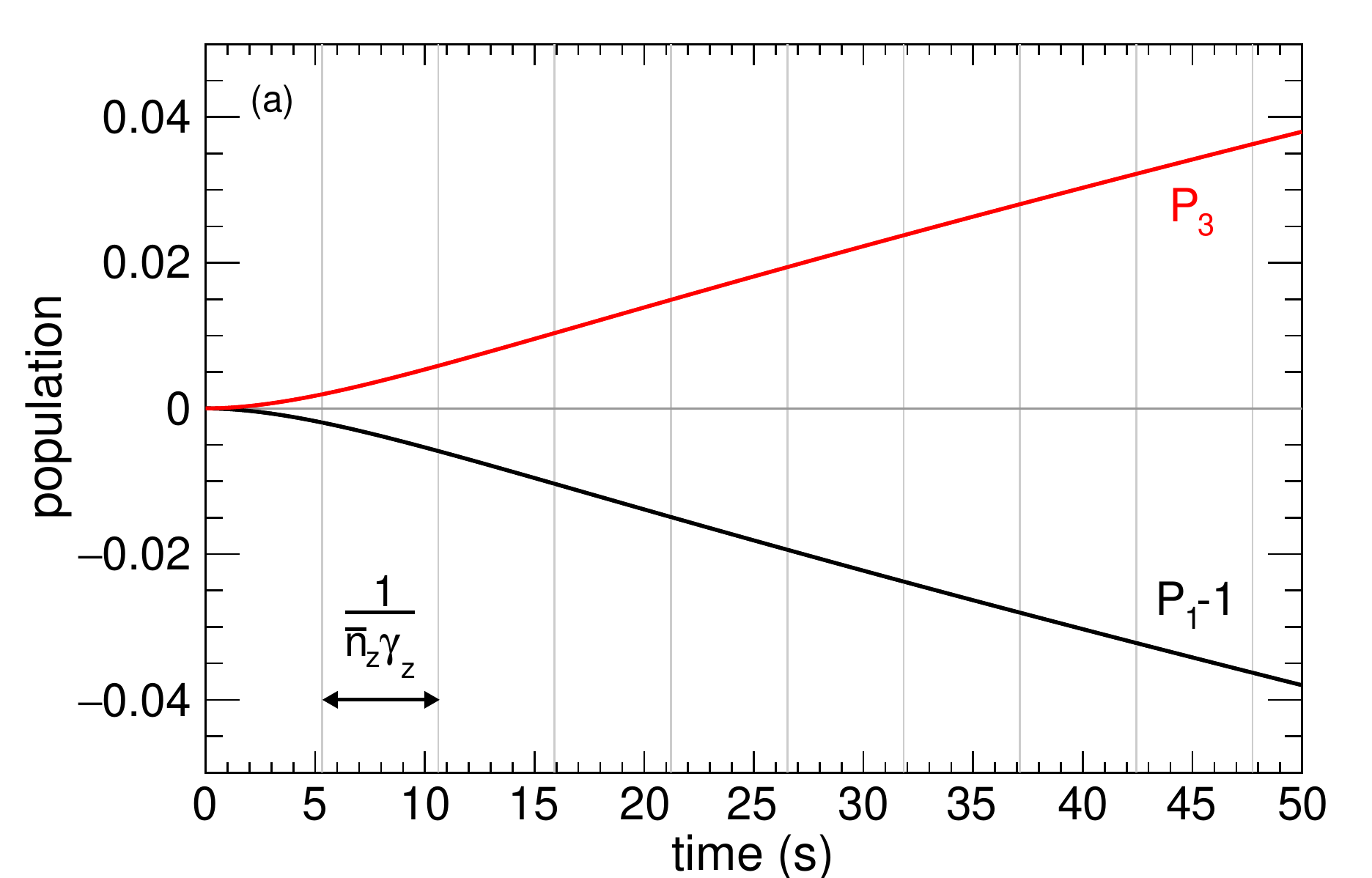}
     \includegraphics[width=\the\columnwidth]{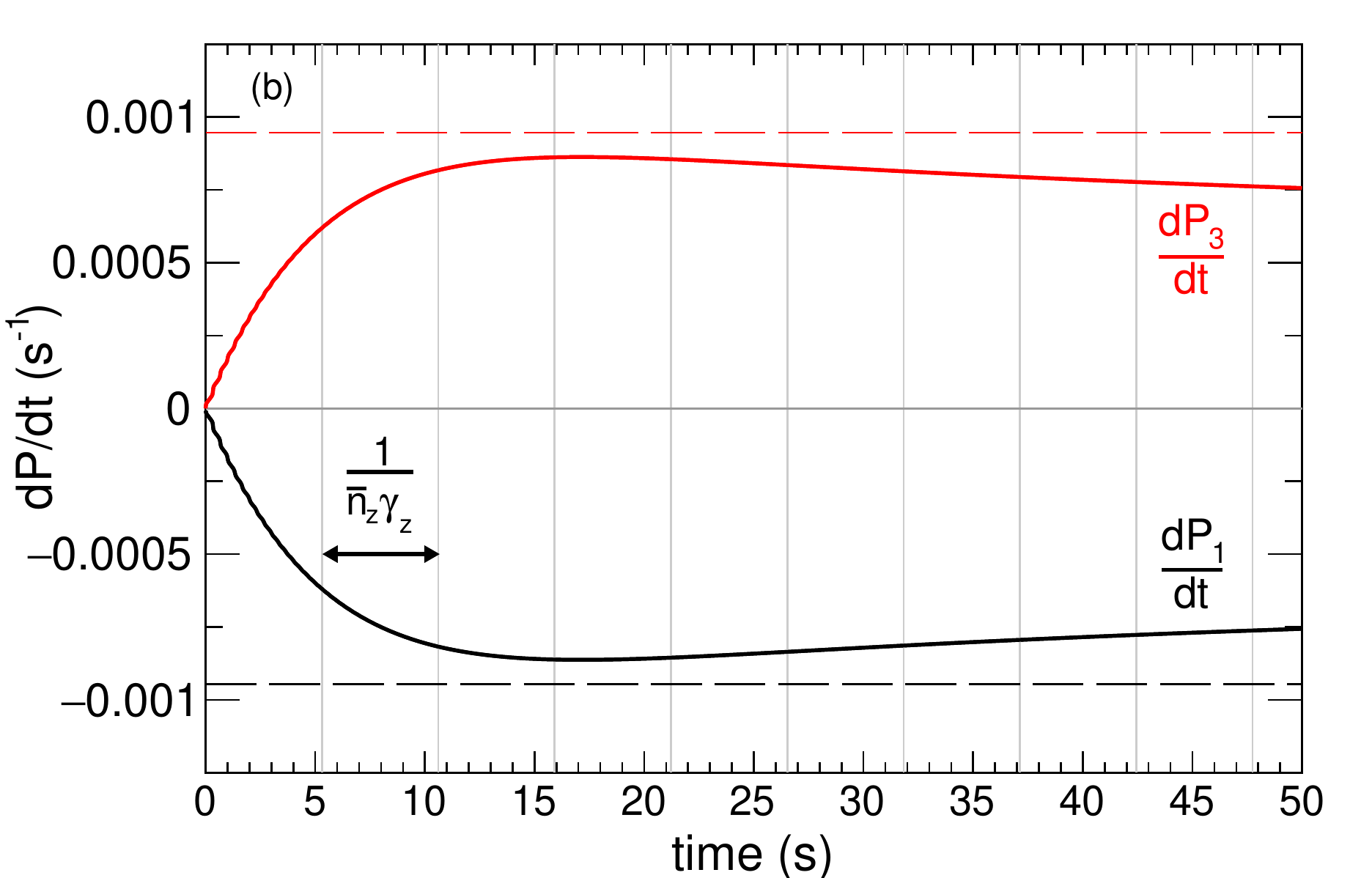}
        \caption{The probabilities for spin up (red) and  and spin down minus 1 (black) in (a), and for the derivatives of these probabilities (b) as a function of time for a spin flip drive tuned to the $n_z=0$ cyclotron resonance, with Rabi frequency $\Omega_s/(2\pi) = 0.01$ Hz and other parameters in Tab~\ref{table:FrequenciesCompared}.  The horizontal grids are spaced by $(\bar{n}_z \gamma_z)^{-1}$  The vertical grids show quasi steady state values. 
    }    
    \label{Fig:SpinFlipLineShape}
\end{figure}

Eq.~(\ref{eq:MatrixEquationSpinC}) gives 
\begin{equation}
\frac{d}{dt}\vec{p}_{33}(t)=\Omega_s\textrm{Im}\left[\vec{p}_{13}(t)\right], 
\end{equation}
when a small excitation $\vec{p}_{33} \approx 0$ is assumed. The steady-state $\vec{p}_{13}(t)$ comes from solving Eq.~(\ref{eq:MatrixEquationSpinB}),  
\begin{equation}
\begin{split}
\vec{p}_{13}(t)
&=\mathbf{R}^{-1}(\epsilon_s,\delta_s,0)\left[-i\frac{\Omega_s}{2}(\vec{p}_{33}(t)-\vec{p}_{11}(t))\right]\\
&=i\frac{\Omega_s}{2}\mathbf{R}^{-1}(\epsilon_s,\delta_s,0)\, \vec{p}(T),
\end{split}
\end{equation}
for $\vec{p}_{11} \approx \vec{p}(T)$ and $\vert\vec{p}_{33}\vert \ll 1$. The transition rate is then  
\begin{equation}
\frac{dP_3}{dt} =\frac{\Omega_s^2}{2}\mathrm{Im}\left[\sum_{{n_z}=0}^{\infty}\left(i\mathbf{R}^{-1}(\epsilon_s,\delta_s,0) \vec{p}(T)\right)_{n_z}\right].
\label{eq:TransitionRateSpin}
\end{equation}
This rate is essentially the line shape defined in Eq.~(\ref{eq:P}) except for the cyclotron damping rate $\gamma_c$ and the parameters for spin flip transition $\Omega_s$, $\epsilon_s$ and $\delta_s$.

\begin{figure}
    \centering
    \includegraphics[width=\the\columnwidth]{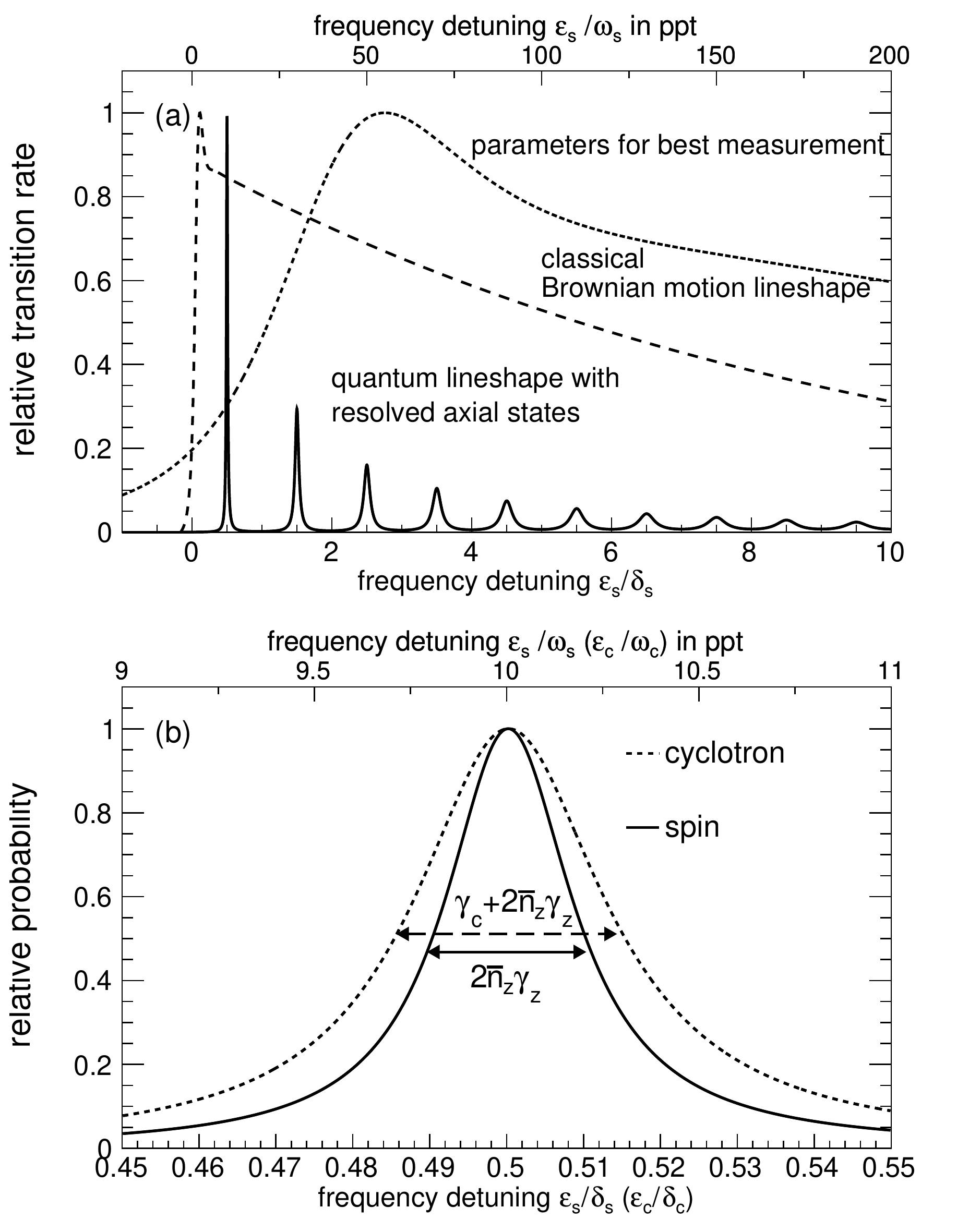}
        \caption{(a) Quantum  spin-flip line shape  (solid) for a weak drive with the largest peak normalized to 1 and for the classical Brownian motion line shape(dashed). The line shape used for the best measurement (dotted) is also shown. (b) Normalized probability of a spin-flip transition (solid) compared to the cyclotron transition (dashed). The horizontal scale is $\epsilon_s/\delta_s$ for the spin transition and $\epsilon_c/\delta_c$ for the cyclotron transition.
    }    
    \label{Fig:SpinFlipLineShape}
\end{figure}

The directly driven spin flip transition rate in Fig.~(\ref{Fig:SpinFlipLineShape}a) is very similar to the cyclotron line shape of Fig.~(\ref{fig:CyclotronLineshape}a) for same experimental conditions in Table~\ref{table:FrequenciesCompared}.  A spin-flip resonance for every axial quantum state is clearly resolved. The line shapes for classical calculation (Eq.~(\ref{eq:classicallineshape1})) and for the best measurement parameters (Tab.~\ref{table:FrequenciesComparedOld}) are also shown for comparison. Figure~\ref{Fig:SpinFlipLineShape}(b) compares the line shape for spin flip (solid line) compared to the one for the cyclotron transition (dashed line) with the parameters in Tab.~\ref{table:FrequenciesCompared}. Since the magnetic bottle parameters are related by $\delta_s=g/2\times\delta_c\approx1.001\delta_c$, the transition rate line shape for spin flip (Eq.~(\ref{eq:TransitionRateSpin})) is much the same as the cyclotron line shape. The only difference from the small damping rate $\gamma_s\approx0$ appears when focusing on the $n_z=0$ peak (fig.~\ref{Fig:SpinFlipLineShape}(b)). Because of the negligible spin-flip damping rate, the full-width at the half maximum of the peak $2\bar{n}_z\gamma_z$ is slightly narrower than the cyclotron's linewidth $\gamma_c+2\bar{n}_z\gamma_z$. The spin transition line shape peaks can be made even narrower by reducing $\gamma_z$ further. 
The possibility to use the spin-flip transition probability is discussed and compared to alternatives in Sec.~\ref{sec:Comparison}.

\section{Two-drive Spin Flips}
\label{sec:twophotoncalculation}

\subsection{Master Equation}

Spin flips (from the spin down ground state $\left|1\right\rangle$ to the spin up ground state $\left|3\right\rangle$) can also be driven using simultaneously applied cyclotron and anomaly drives ($\Omega_c > 0$ and $\Omega_a > 0$ with $\Omega_s=0$) instead of the direct spin flip drive  discussed in Sec.~\ref{sec:DirectSpinFlips}.  A practical advantage is that the stable final state would remain unchanged as long as is needed to detect it.  This was true for the anomaly transitions considered above, but not for cyclotron transitions that must be detected before cyclotron decay. A quantum calculation is needed to ascertain whether two photon transitions would be less sensitive to slow drifts of the magnetic field insofar as the spin and cyclotron motion will experience the same average magnetic field.  

The three sets of states in Eq.~(\ref{eq:ThreeStates}) are involved in flipping the spin via the two drives.  The density operator in the interaction picture can be written in terms of operators $p_{jk}$ proportional to the operators $\tilde{\rho}_{jk}$ of Eq.~(\ref{eq:rhojk}), such that 
\begin{equation}
\begin{split}
\tilde{\rho} =& 
\begin{pmatrix} 
p_{11} & p_{12} e^{i\epsilon_c t} & p_{13} e^{i(\epsilon_c+\epsilon_a)t} \\ 
p_{21} e^{-i\epsilon_c t} & p_{22} & p_{23} e^{i \epsilon_a t} \\ 
p_{31} e^{-i(\epsilon_c+\epsilon_a)t} & p_{32} e^{-i \epsilon_a t} & p_{33}
\end{pmatrix}.
\end{split}
\end{equation}
The transformation puts the master equation in the form
\begin{equation}
\begin{split}
\frac{d}{dt} &
\begin{pmatrix} 
p_{11} & p_{12} & p_{13} \\
p_{21} & p_{22} & p_{23}\\
p_{31} & p_{32} & p_{33}
\end{pmatrix}
=\\
&-i
\begin{pmatrix} 
0                                       & \epsilon_c p_{12}   & (\epsilon_c+\epsilon_a)p_{13} \\
-\epsilon_c p_{21}                      & 0                  &\epsilon_a p_{23}\\
-(\epsilon_c+\epsilon_a)p_{31}          & -\epsilon_a p_{32} & 0
\end{pmatrix} \\
& -{i}\left(a_z^{\dagger}a_z+\tfrac{1}{2}\right)
\begin{pmatrix} 
0              & -\delta_c p_{12} &-\delta_s p_{13}\\
\delta_c p_{21} & 0               & -\delta_a p_{23}\\
\delta_s p_{31} & \delta_a p_{32}  & 0
\end{pmatrix}\\
&-\frac{i\Omega_c}{2} 
\begin{pmatrix} 
i2 \mathrm{Im}[ p_{21}]     & p_{22}- p_{11}             &  p_{23}\\
 p_{11}- p_{22}             &i2 \mathrm{Im}[p_{12}]      &  p_{13}\\
-p_{32}                     &-p_{31}                     &0
\end{pmatrix}\\
& -\frac{i\Omega_a}{2}
\begin{pmatrix} 
0          &-p_{13}                      &-p_{12}\\
p_{31}     &i2  \mathrm{Im}[ p_{32}]     &  p_{33}- p_{22}\\
p_{21}     &p_{22}- p_{33}               &i2  \mathrm{Im}[p_{23}]
\end{pmatrix}\\
&-\frac{\gamma_c}{2}
\begin{pmatrix} 
-2p_{22}      & p_{12}      & 0     \\
p_{21}        & 2p_{22}     & p_{23}\\
0             & p_{32}      & 0
\end{pmatrix}\\
&-\frac{\gamma_z}{2}\bar{n}_z\left(a_z a_z^\dagger p -2a_z^\dagger p a_z+p a_z a_z^\dagger\right)\\
&-\frac{\gamma_z}{2}\left(\bar{n}_z+1\right)\left(a_z^\dagger a_z p-2a_z p a_z^\dagger+p a_z^\dagger a_z\right).
\end{split}
\label{eq:eqtwophotonp}
\end{equation}
All time dependence is now within the components $p_{ij}$.  

For a thermal distribution of initial axial states, the operators $p_{ij}$ are axially diagonal, with diagonal components, $p_{ij;n_z}=\langle i,n_z|p_{ij}|j,n_z\rangle$.  The master equation is then given by the differential equations,
\begin{subequations}
\begin{align}
\frac{d}{dt}&p_{11;n_z}(t)\nonumber\\
&=\left[-\gamma_z(2\bar{n}_z+1)n_z-\gamma_z\bar{n}_z\right]p_{11;n_z}(t)\nonumber\\
&+\gamma_z\bar{n}_zn_zp_{11;n_z-1}(t)+\gamma_z(\bar{n}_z+1)(n_z+1)p_{11;n_z+1}(t)\nonumber\\
&-\Omega_c\textrm{Im}\left[p_{12;n_z}(t)\right]+\gamma_cp_{22;n_z}(t)\\
\frac{d}{dt}&p_{22;n_z}(t)\nonumber\\
&=\left[-\gamma_c-\gamma_z(2\bar{n}_z+1)n_z-\gamma_z\bar{n}_z\right]p_{22;n_z}(t)\nonumber\\
&+\gamma_z\bar{n}_zn_zp_{22;n_z-1}(t)+\gamma_z(\bar{n}_z+1)(n_z+1)p_{22;n_z+1}(t)\nonumber\\
&+\Omega_c\textrm{Im}\left[p_{12;n_z}(t)\right]-\Omega_a\textrm{Im}\left[p_{23;n_z}(t)\right]\\
\frac{d}{dt}&p_{33;n_z}(t)\nonumber\\
&=\left[-\gamma_z(2\bar{n}_z+1)n_z-\gamma_z\bar{n}_z\right]p_{33;n_z}(t)\nonumber\\
&+\gamma_z\bar{n}_zn_zp_{33;n_z-1}(t)+\gamma_z(\bar{n}_z+1)(n_z+1)p_{33;n_z+1}(t)\nonumber\\
&+\Omega_a\textrm{Im}\left[p_{23;n_z}(t)\right]\\
\frac{d}{dt}&p_{12;n_z}(t)\nonumber\\
&=\big[i\left(-\epsilon_c+\delta_c\left(n_z+\tfrac{1}{2}\right)\right)\nonumber\\
&-\tfrac{1}{2}\gamma_c-\gamma_z(2\bar{n}_z+1)n_z-\gamma_z\bar{n}_z\big]p_{12;n_z}(t)\nonumber\\
&+\gamma_z\bar{n}_zn_zp_{12;n_z-1}(t)+\gamma_z(\bar{n}_z+1)(n_z+1)p_{12;n_z+1}(t)\nonumber\\
&-i\frac{\Omega_c}{2}(p_{22;n_z}(t)-p_{11;n_z}(t))+i\frac{\Omega_a}{2}p_{13;n_z}(t)\\
\frac{d}{dt}&p_{23;n_z}(t)\nonumber\\
&=\big[i\left(-\epsilon_a+\delta_a\left(n_z+\tfrac{1}{2}\right)\right)\nonumber\\
&-\tfrac{1}{2}\gamma_c-\gamma_z(2\bar{n}_z+1)n_z-\gamma_z\bar{n}_z\big]p_{23;n_z}(t)\nonumber\\
&+\gamma_z\bar{n}_zn_zp_{23;n_z-1}(t)+\gamma_z(\bar{n}_z+1)(n_z+1)p_{23;n_z+1}(t)\nonumber\\
&-i\frac{\Omega_a}{2}(p_{33;n_z}(t)-p_{22;n_z}(t))-i\frac{\Omega_c}{2}p_{13;n_z}(t)\\
\frac{d}{dt}&p_{13;n_z}(t)\nonumber\\
&=\big[i\left(-\left(\epsilon_c+\epsilon_a\right)+\delta_s\left(n_z+\tfrac{1}{2}\right)\right)\nonumber\\
&-\gamma_z(2\bar{n}_z+1)n_z-\gamma_z\bar{n}_z\big]p_{13;n_z}(t)\nonumber\\
&+\gamma_z\bar{n}_zn_zp_{13;n_z-1}(t)+\gamma_z(\bar{n}_z+1)(n_z+1)p_{13;n_z+1}(t)\nonumber\\
&-i\frac{\Omega_c}{2}p_{23;n_z}(t)+i\frac{\Omega_a}{2}p_{12;n_z}(t).
\end{align}    
\end{subequations}
The general time-dependent solution  of these equations has initial conditions  $p_{11,n_z}(0) = p_{n_z}(T)$, with $p_{ij,n_z}(0)=0$ all other $i$ and $j$.  

In terms of  $\mathbf{R}(\epsilon,\delta,\gamma_c)$ from Eq.~(\ref{eq:R}) and vectors $\vec{p}_{ij}$ with components $p_{ij;n_z}=\langle i,n_z|p_{ij}|j,n_z\rangle$, the vector equations of motion are
\begin{subequations}
\begin{align}
\frac{d}{dt}\vec{p}_{11}(t)
&=\mathbf{R}(0,0, 0)\vec{p}_{11}(t) \nonumber \\
&-\Omega_c\mathrm{Im}[\vec{p}_{12}(t)]+\gamma_c\vec{p}_{22}(t)\\
\frac{d}{dt}\vec{p}_{22}(t)&=\mathbf{R}(0,0, 2\gamma_c)\vec{p}_{22}(t)\nonumber \\
&+\Omega_c\mathrm{Im}[\vec{p}_{12}(t)]-\Omega_a\mathrm{Im}[\vec{p}_{23}(t)]\\
\frac{d}{dt}\vec{p}_{33}(t)&=\mathbf{R}(0,0,0)\vec{p}_{33}(t)\nonumber \\
&+\Omega_a\mathrm{Im}[\vec{p}_{23}(t)]\\
\frac{d}{dt}\vec{p}_{12}(t)&=\mathbf{R}(\epsilon_c,\delta_c,\gamma_c)\vec{p}_{12}(t)\nonumber \\
&-i\tfrac{\Omega_c}{2}\left(\vec{p}_{22}(t)-\vec{p}_{11}(t)\right)+i\tfrac{\Omega_a}{2}\vec{p}_{13}(t)\\
\frac{d}{dt}\vec{p}_{23}(t)&=\mathbf{R}(\epsilon_a,\delta_a, \gamma_c)\vec{p}_{23}(t)\nonumber \\
&-i\tfrac{\Omega_a}{2}\left(\vec{p}_{33}(t)-\vec{p}_{22}(t)\right)-i\tfrac{\Omega_c}{2}\vec{p}_{13}(t)\\
\frac{d}{dt}\vec{p}_{13}(t)&=\mathbf{R}(\epsilon_c+\epsilon_a,\delta_s,0)\vec{p}_{13}(t)\nonumber \\
&-i\tfrac{\Omega_c}{2}\vec{p}_{23}(t)+i\tfrac{\Omega_a}{2}\vec{p}_{12}(t).
\end{align}
\label{eq:TwoPhotonVector}%
\end{subequations}
The initial conditions are  $\vec{p}_{11}(0)=\vec{p}(T)$, with $\vec{p}_{ij}(0)=0$ for all other $i$ and $j$. Small, non-resonant  excitations to more highly excited states are neglected.   For the parameters being considered in this work, we found that simultaneously solving 900 differential equations determine the solution to the master equation numerically for cyclotron and anomaly drives applied at the same time.  

\subsection{Quasi Steady State}

A quasi steady state is produced when weak cyclotron and anomaly drives, with
\begin{eqnarray}
  &\Omega_c \ll \gamma_c\\
  &\Omega_a \ll \gamma_c, 
\end{eqnarray}
are applied for a time $t$ in the range
\begin{equation}
 \gamma_c^{-1}\ll t \ll \gamma_c/\Omega_a^2.  
\end{equation} 
The time must be long compared to the cyclotron damping time to allow transients to dies out.  It must be short compared to the time it takes to transfer an appreciable population to the spin up spin states.  

\begin{figure}[htbp!]
    \centering
    \includegraphics[width=\the\columnwidth]{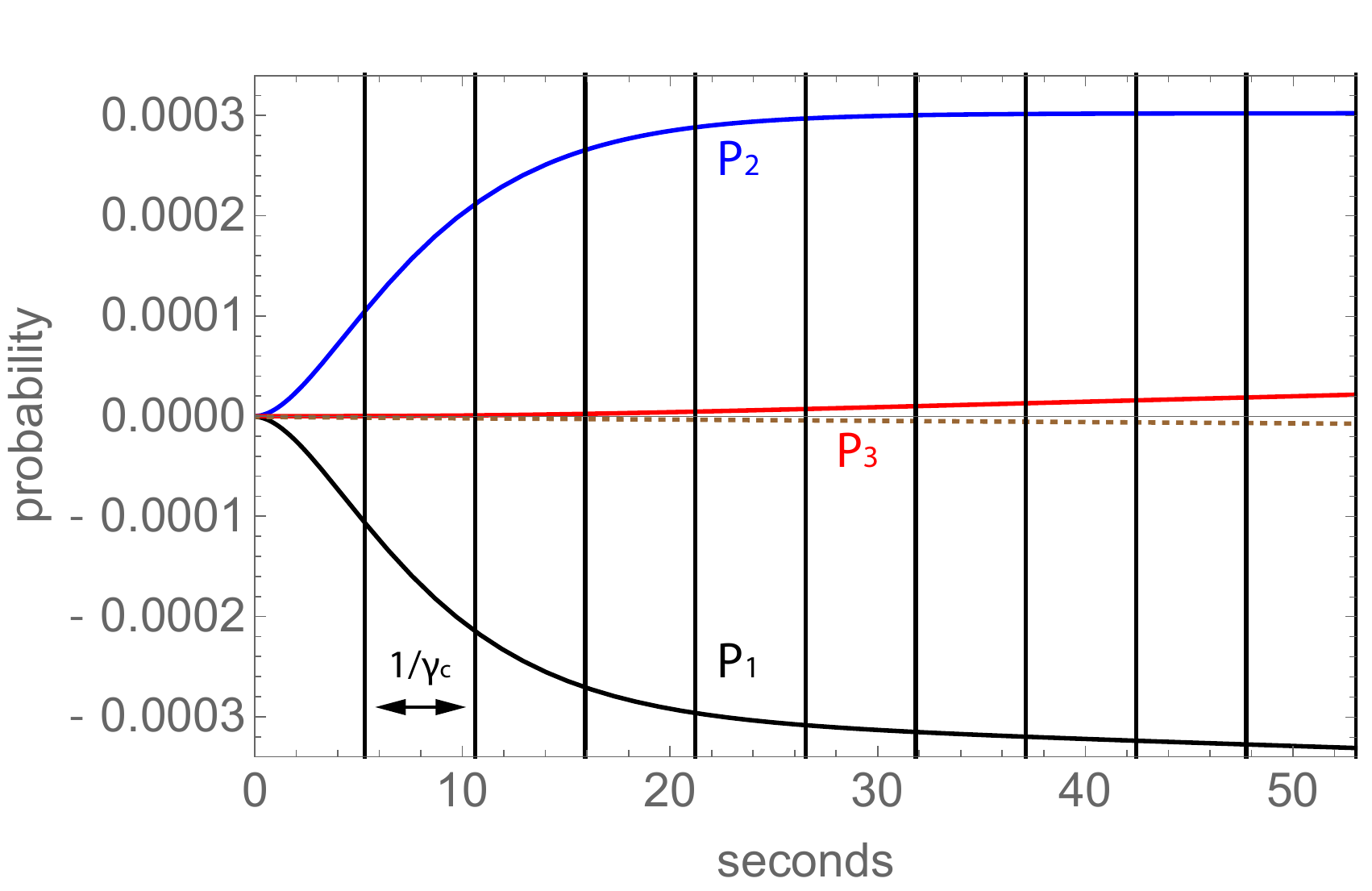}
    \caption{Time evolution driven by weak and resonant cyclotron and anomaly drives applied for 10 cyclotron damping times, the latter indicated by vertical gray lines. The probability to be in the $\left|2,n_z\right\rangle$ states (blue) reaches a steady state after transients die out on a time scale give by the cyclotron damping time, $1/\gamma_c$. The  probability to be in the initial $\left|1,n_z\right\rangle$ states, with  unit probability subtracted out, is shown in black. The probability to be be driven to the final $\left|3,n_z\right\rangle$ states is shown in red.}    \label{fig:SimultaneousTimeEvolution}
\end{figure}

\begin{figure}[htbp!]
    \centering
    \includegraphics[width=\the\columnwidth]{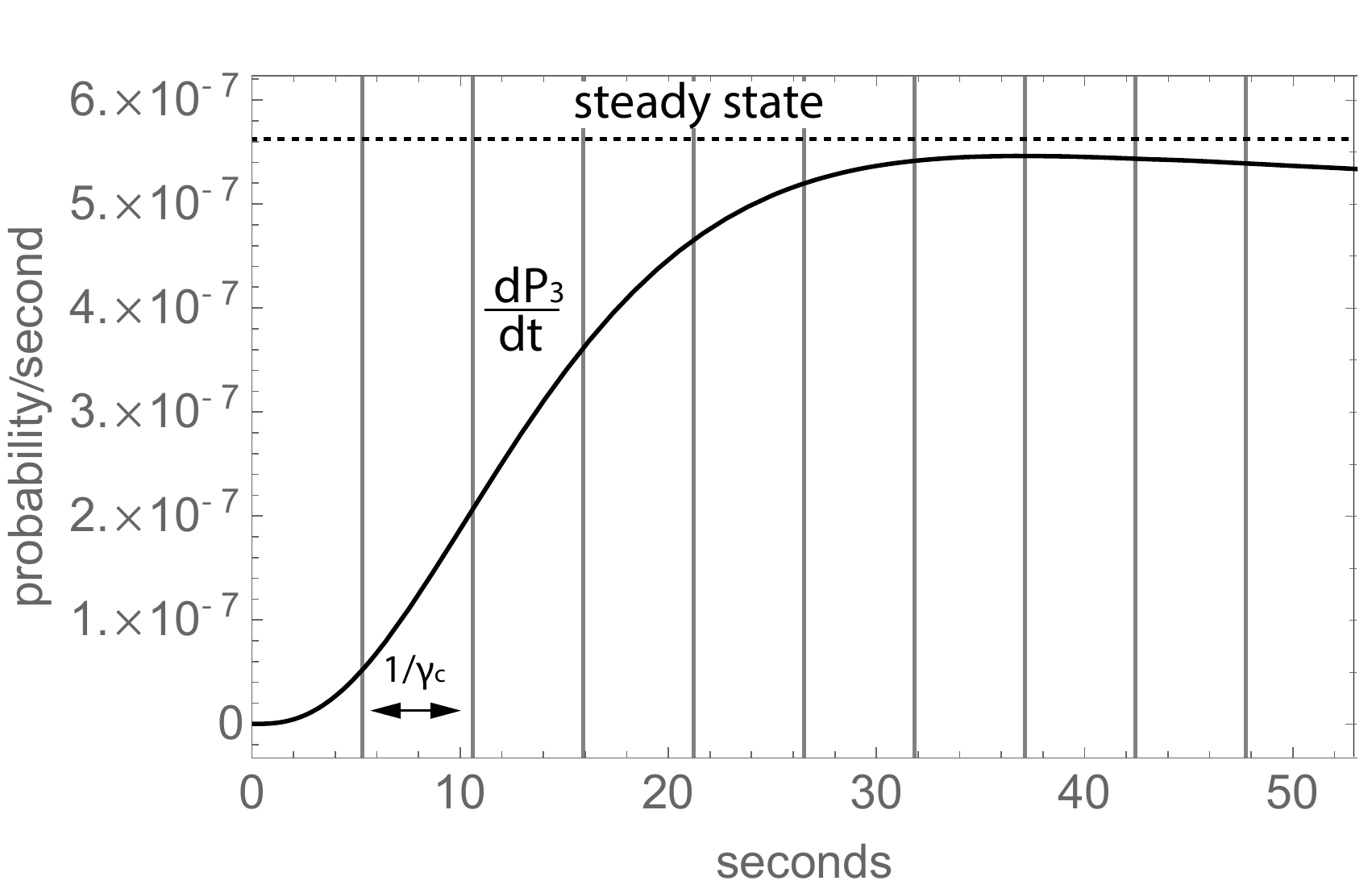}
    \caption{Rate $dP_3/dt$ for simultaneously applied weak cyclotron and anomaly drives ($\Omega_c = \Omega_a=\gamma_c/10$) that are resonant. The quasi steady state estimate (dashed) slightly overstates the transition rate.
    }    
    \label{fig:SimultaneousD3}
\end{figure}

 Fig.~\ref{fig:SimultaneousTimeEvolution} illustrates the time evolution for weak drives ($\Omega_c=\Omega_a = 0.1 \gamma_c$ that are resonant, and for the realistic experimental conditions in Table~\ref{table:FrequenciesCompared}. The sum of the probabilities to be in the states $\left|l,n_z\right\rangle$ 
 \begin{equation}
     P_l = \sum_{n_z=0}^{\infty} p_{ll;n_z}
 \end{equation}
 (from Eqs.~(\ref{eq:Pl}) and (\ref{eq:Plp})) is plotted for $l=1,2,3$.    The drives are turned on at time  $t=0$ and the time evolution shown continues for ten cyclotron damping period, to $t=10/\gamma_c$.  The probability $P_2$ to be driven into the  $\left|2,n_z\right\rangle$ states (blue) increases from zero to reach a quasi steady state after the transients die out in several cyclotron damping times $1/\gamma_c$.  The probability $P_1$ to remain in the initial  $\left|1,n_z\right\rangle$ states, minus unit probability, is shown in black.  It remains at essentially unit probability, decreasing only slightly to conserve probability.  
 The much smaller probability (solid red) to transition to the cyclotron ground states with spin up, $\left|3,n_z\right\rangle$, gradually increases at first, and then increases linearly for much of the $10/ \gamma_c$ time evolution.  The solid red curve in Fig.~\ref{fig:SimultaneousD3} illustrates how it is the derivative $dP_3/dt$ that reaches a quasi steady state.  
 
An approximate analytic expression for the quasi steady state rate \begin{equation}
\frac{dP_3}{dt}=\frac{d}{dt} \sum_{n_z=0}^\infty p_{33;n_z}(t)=
\Omega_a\mathrm{Im}\left[\sum^\infty_{n_z=0}{p}_{23;n_z}(t)\right],
\label{eq:E6}
\end{equation}
comes from summing Eq.~(\ref{eq:TwoPhotonVector}c) over all axial states and simplifying using $\sum^\infty_{n_z=0}\left(\mathbf{R}(0,0,0)\vec{p}_{33}\right)_{n_z}=0$. The quasi steady state is also described by
\begin{subequations}
\begin{align}
&\vec{p}_{11} = \vec{p}(T) \label{eq:E1}\\
&\mathbf{R}(0,0, 2\gamma_c)\vec{p}_{22}+\Omega_c\mathrm{Im}[\vec{p}_{12}]=0\label{eq:E2}\\
&\mathbf{R}(\epsilon_c,\delta_c,\gamma_c)\vec{p}_{12}+i\tfrac{\Omega_c}{2}\vec{p}_{11}=0\label{eq:E3}\\
&\mathbf{R}(\epsilon_a,\delta_a, \gamma_c)\vec{p}_{23} -i\tfrac{\Omega_a}{2}(\vec{p}_{33}-\vec{p}_{22})-i\tfrac{\Omega_c}{2}\vec{p}_{13}=0\label{eq:E4}\\
&\mathbf{R}(\epsilon_c+\epsilon_a,\delta_s,0)\vec{p}_{13}+i\tfrac{\Omega_a}{2}\vec{p}_{12}=0.
\label{eq:E5}
\end{align}
\end{subequations}
The first of these equations states that $\vec{p}_{11}$ then remains at the initial thermal equilibrium value, $\vec{p}_{n_z}(T)$.  The remaining equations assume  $\vert{p}_{11;n_z}\vert\gg\vert{p}_{22;n_z}\vert$, $\vert{p}_{11;n_z}\vert\gg\vert{p}_{13;n_z}\vert$, and $\vert{p}_{12;n_z}\vert\gg\vert{p}_{32;n_z}\vert$, and the time derivatives of $\vec{p}_{12}$, $\vec{p}_{22}$ and $\vec{p}_{23}$ are neglected. The drives must also be applied for a time $t \gg (\bar{n}_z\gamma_z)^{-1}$, so that the 
time derivative of $\vec{p}_{13}$ can be neglected because of the decoherence of the superposition of axial states at a rate $\bar{n}_z\gamma_z$. (This last condition does not hold when there in only one axial state in the  $T \rightarrow 0$ limit.)  

The solutions to these linear equations are
\begin{subequations}
\begin{align}
\vec{p}_{12}&=-i\tfrac{\Omega_c}{2}\mathbf{R}(\epsilon_c,\delta_c,\gamma_c)^{-1}\vec{p}(T)\label{eq:eqtwophotonPLinear1}
\\
\vec{p}_{22}&=-\Omega_c\mathbf{R}(0,0,2\gamma_c)^{-1}\mathrm{Im}[\vec{p}_{12}]\label{eq:eqtwophotonPLinear2}
\\
\vec{p}_{13}&=-i\tfrac{\Omega_a}{2}\mathbf{R}(\epsilon_c+\epsilon_a,\delta_s,0)^{-1}\vec{p}_{12}\label{eq:eqtwophotonPLinear3}
\\
\vec{p}_{23}&=\tfrac{1}{2}\mathbf{R}(\epsilon_a,\delta_a, \gamma_c)^{-1}
\left[i\Omega_c\vec{p}_{13}-i\Omega_a\vec{p}_{22}+i\Omega_a\vec{p}_{33}\right].
\label{eq:eqtwophotonPLinear4}
\end{align}
\end{subequations}
The last of these equations does not yet describe a steady state because it depends upon the growing $\vec{p}_{33}$. 

Eq.~(\ref{eq:eqtwophotonPLinear4}) can be substituted into  Eq.~(\ref{eq:E6}) to allow an estimate of how rapidly $\vec{p}_{33}$ grows in time. The time dependent parts are 
\begin{equation}
\begin{split}
\frac{d}{dt} \sum_{n_z=0}^{\infty} {p}_{33;n_z}
&=\frac{\Omega_a^2}{2}\mathrm{Im}\left[i\sum^\infty_{n_z}\left(\mathbf{R}(\epsilon_a,\delta_a, \gamma_c)^{-1} \vec{p}_{33}\right)_{n_z}\right]+\mathcal{C},
\end{split}
\label{eq:p33timeconstant}
\end{equation}
with $\mathcal{C}$ represents terms which do not depend upon time for $t\gg\gamma_c^{-1}$.  Roughly speaking, $\vec{p}_{33}$ (i.e.\ the diagonal elements of $p_{33}$) approaches its steady state with a rate going as  $\Omega_a^2|\mathbf{R}(\epsilon_a,\delta_a, \gamma_c)^{-1}|$.  The magnitude of the transformation matrix roughly goes as its eigenvalues, $\bar{n}_z\gamma_z + \gamma_c>\gamma_c$.  This means that the  time constant is longer than $\Omega_a^2/\gamma_c$, more than a thousand seconds with realistic parameters in Table~\ref{table:FrequenciesCompared}. For a realistic drive time  $t\ll(\Omega_a^2/\gamma_c)^{-1}$, the last term in Eq.~(\ref{eq:eqtwophotonPLinear4}) can be neglected, as needed to make a steady state equation.   

The quasi steady state spin-flip rate as a function of detunings ${dP_3(\epsilon_c,\epsilon_a)}/{dt}$ is thus
\begin{subequations}
\begin{align}
\frac{dP_3(\epsilon_c,\epsilon_a)}{dt}
&=\frac{\Omega_a^2\Omega_c^2}{8}\textrm{Im} 
\Bigg[
\sum^\infty_{n_z=0}
\bigg(i \mathbf{R}(\epsilon_a,\delta_a,\gamma_c)^{-1}  \vec{W} \bigg)_{n_z} \Bigg],\label{eq:twophototransitionA}    
\\
\vec{W} = &-2\mathbf{R}(0,0,2\gamma_c)^{-1}\mathrm{Im}[i\mathbf{R}(\epsilon_c,\delta_c,\gamma_c)^{-1}\vec{p}(T)]\nonumber\\
&-\mathbf{R}(\epsilon_c+\epsilon_a,\delta_s,0)^{-1}\mathbf{R}(\epsilon_c,\delta_c,\gamma_c)^{-1}\vec{p}(T)
\label{eq:twophototransitionB}    
\end{align}
\end{subequations}
using Eqs.(\ref{eq:E6}) and (\ref{eq:eqtwophotonPLinear1}-\ref{eq:eqtwophotonPLinear4}). The first term in $\vec{W}$ describes sequential one photon transitions (Fig.~\ref{fig:SimultaneousContour}(b)).  The second term in $\vec{W}$, depending as it does upon $ \mathbf{R}(\epsilon_c+\epsilon_a,\delta_s,0)$, adds the effect of direct two photon transitions (Fig.~\ref{fig:SimultaneousContour}(c)).

Figure~\ref{fig:SimultaneousD3} compares this quasi steady state derivative (dashed) from the complete solution (solid).  The derivative rises to almost the quasi steady state value and then begins to decrease.

\subsection{Line shapes for Simultaneous Cyclotron and Anomaly Drives}
\label{sec:DisplayedTwoDriveLineshapes}

\begin{figure}[htbp!]
    \centering
    \includegraphics[width=\the\columnwidth]{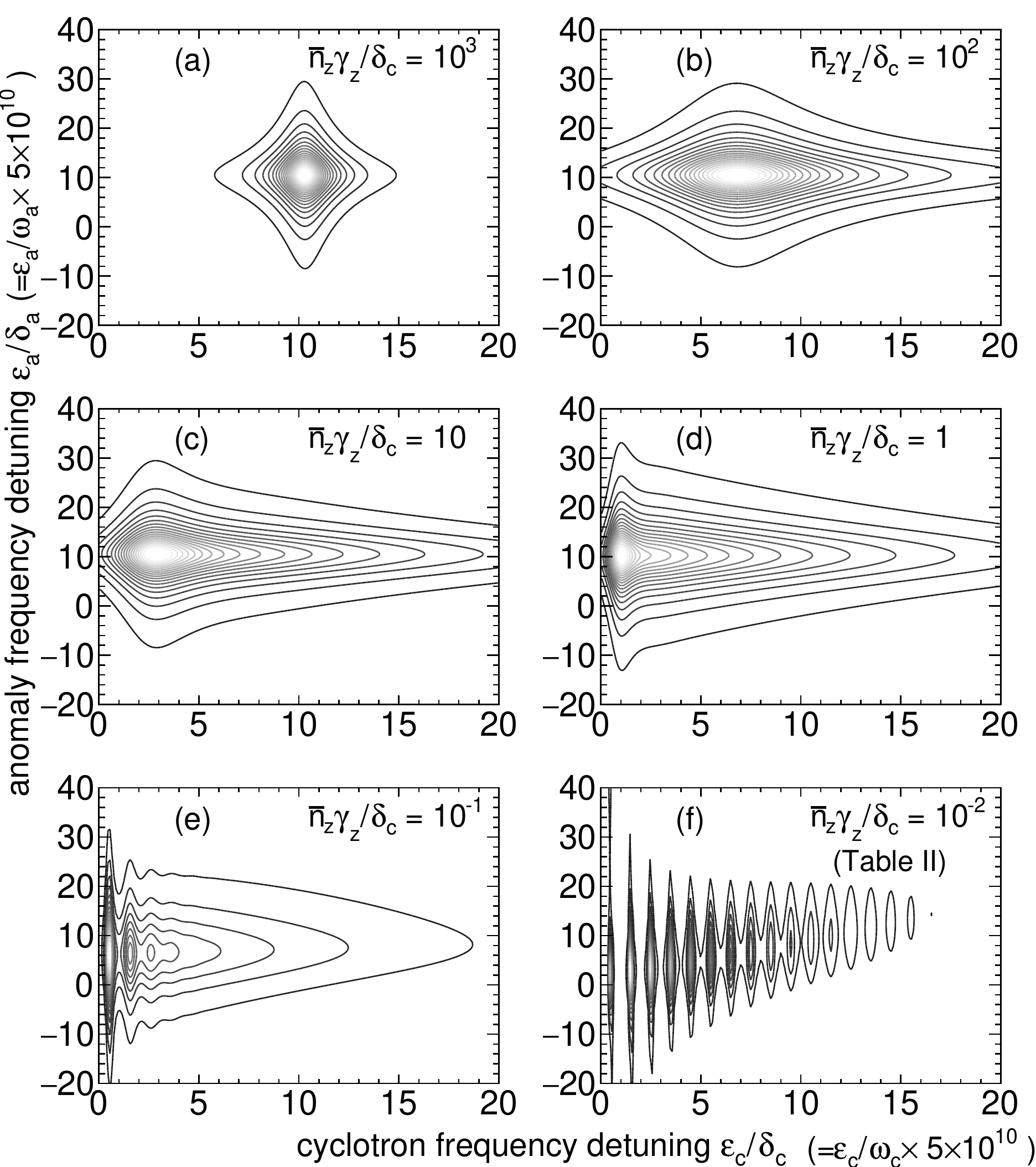}
    \caption{The two-drive line shapes change dramatically as a function of the axial damping rate, $\gamma_z$.  Contours of the quasi steady state ${dP_3(\epsilon_c,\epsilon_a)}/{dt}$ are shown as a function of the scaled detunings from $\omega_c$ and $\omega_a$ of the cyclotron and anomaly drives.  Parameters other than $\gamma_z$ are from  Tab.~\ref{table:FrequenciesCompared}.}
    \label{fig:twophotoweakdrive}
\end{figure}

\begin{figure}[htbp!]
    \centering
    \includegraphics[width=\the\columnwidth]{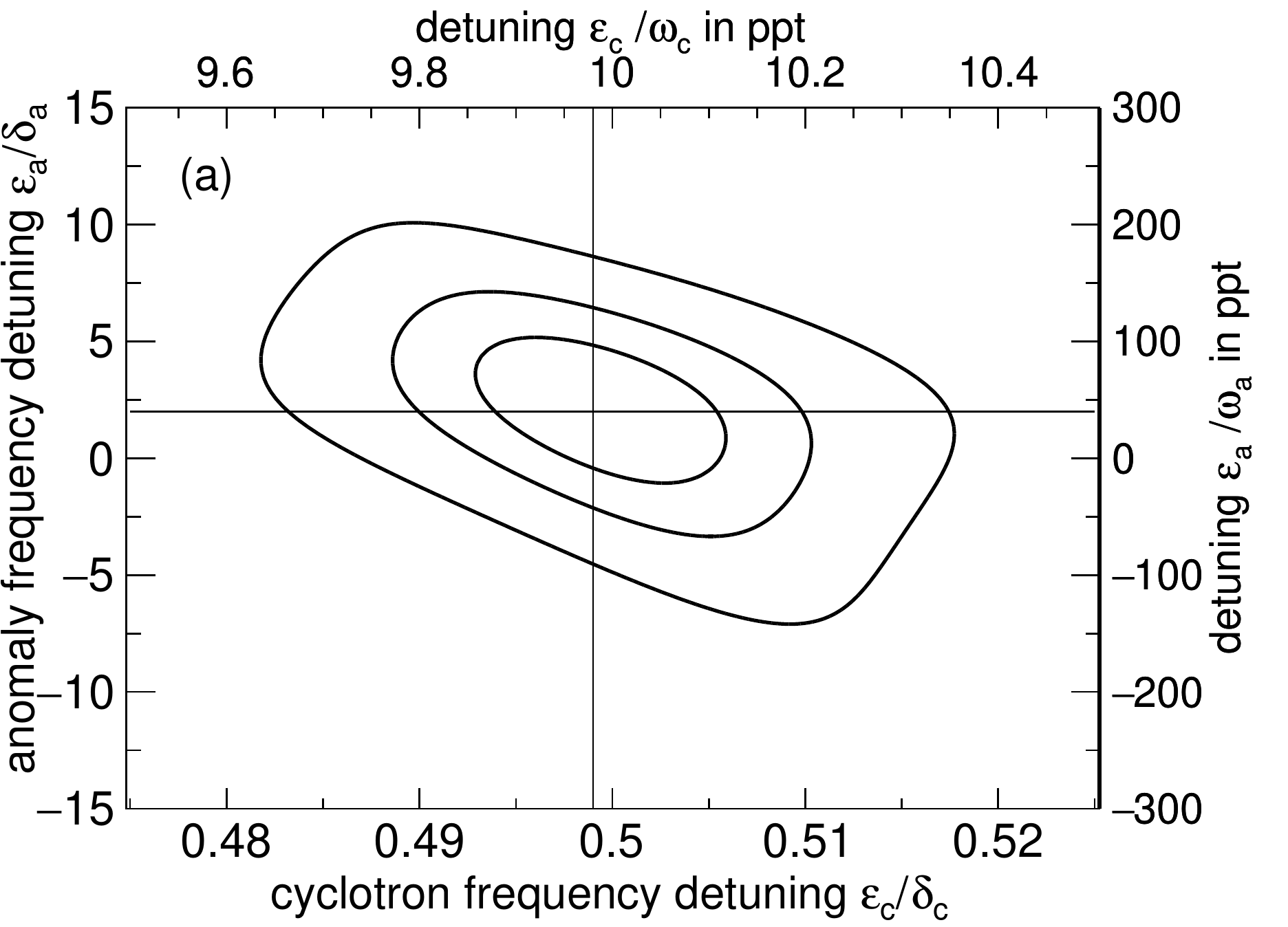}
    \includegraphics[width=\the\columnwidth]{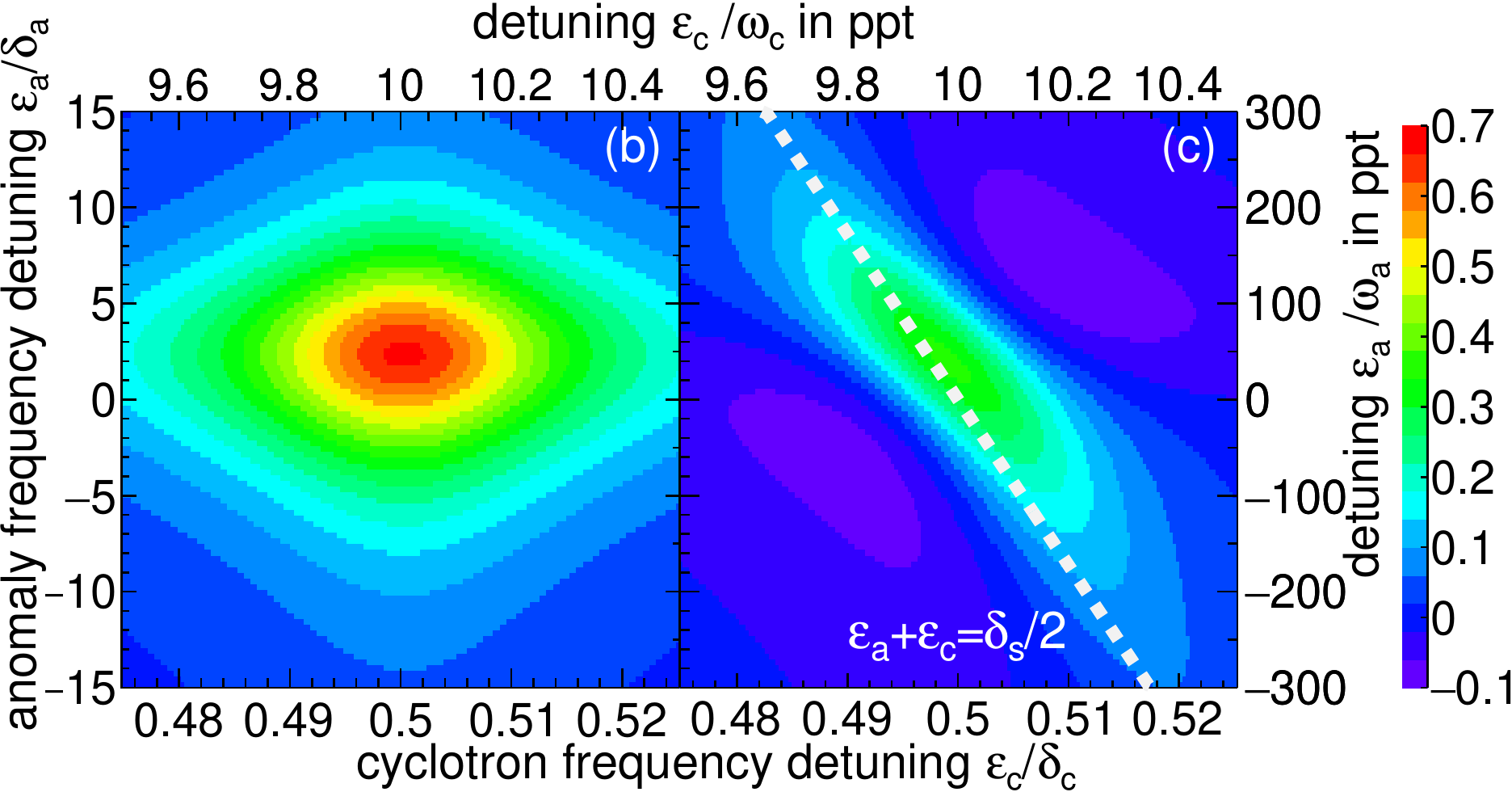}
    \caption{(a) The quasi steady state line shape ${dP_3(\epsilon_c,\epsilon_a)}/{dt}$  for the resolved peak corresponding to $n_z=0$ in Fig.~\ref{fig:twophotoweakdrive}f. The the contours shown are at 75\%, 50\% and 25\% amplitudes relative to the peak value. The anomaly and cyclotron drive frequencies are given in terms of the scaled detunings of these frequencies from $\omega_a$ and $\omega_c$. The interior black lines indicate the drive frequencies scanned in Figs.~\ref{fig:SimultaneousCyclotronLineshape} and \ref{fig:SimultaneousAnomalyLineshape}. (b) and (c) shows the contribution from the first and second terms in Eq.~(\ref{eq:twophototransitionB}) respectively. The colors indicate the amplitude relative to the peak value in (a). The dotted line, $\epsilon_c + \epsilon_a = \delta_s/2$, indicates when the cyclotron and anomaly drive frequencies sum to the resonance $\omega_s+\delta_s/2$. }
    \label{fig:SimultaneousContour}
\end{figure}

Quasi steady state line shapes ${dP_3(\epsilon_c,\epsilon_a)}/{dt}$ are illustrated in 
Figs.~\ref{fig:twophotoweakdrive} and 
\ref{fig:SimultaneousContour} as a function of the anomaly and cyclotron drive frequencies.  The vertical scale is the detuning $\epsilon_a$ of the anomaly drive from $\omega_a$, scaled by $\delta_a$.  The horizontal scale is the detuning $\epsilon_c$ of the cyclotron drive from $\omega_c$, scaled by $\delta_c$. The contours are for probabilities of making a transition from the initial spin-down ground states to spin-up states. 

The dependence of the line shapes upon the axial damping rate $\gamma_z$ is illustrated in Fig.~\ref{fig:twophotoweakdrive}. Except for this damping rate, the experimental parameters from Table~\ref{table:FrequenciesCompared} are used.  As the axial damping rate is lowered, the contributions from individual axial quantum states become resolved as resolved ``islands'' in Fig.~\ref{fig:twophotoweakdrive}f, for the lowest axial damping realized in the laboratory so far \cite{FanRFSwitch2020} while yet allowing quantum jump spectroscopy. 

The narrowest transition peak in Fig.~\ref{fig:twophotoweakdrive}(f), corresponding to $n_z=0$, are potentially the most useful for measuring an electron or positron magnetic moment. Fig.~\ref{fig:SimultaneousContour}(a) shows the contour of 
$dP_3(\epsilon_c,\epsilon_a)/dt$. The contours shown are at 75\%, 50\% and 25\% of the peak amplitude.  The anomaly and cyclotron drive frequencies are specified as scaled detunings of these frequencies from $\omega_a$ and $\omega_c$. Fig.~\ref{fig:SimultaneousContour}(b) and (c) shows the decomposed contributions from the first and second terms in Eq.~(\ref{eq:twophototransitionB}). The dotted line shows where $\epsilon_c + \epsilon_a = \delta_s/2$, which corresponds to the sum of two drive frequencies being equal to $\omega_s+\delta_s/2$. The sum of (b) and (c) gives the tilted contour in (a). 
The drives for this example are weak, with $\Omega_c=\Omega_a=\gamma_c/10$, and the realistic experimental parameters of Table~\ref{table:FrequenciesCompared}) are used. Notice that the peak of the contour slightly deviates from  $\epsilon_c + \epsilon_a = \delta_s/2$.  The anomaly resonance does not resolve into separate peaks for various $n_z$, and the composite peak is thus shifted from $\epsilon_a=\delta_a/2$.

\begin{figure}[htbp!]
  \centering
    \includegraphics[width=\the\columnwidth]{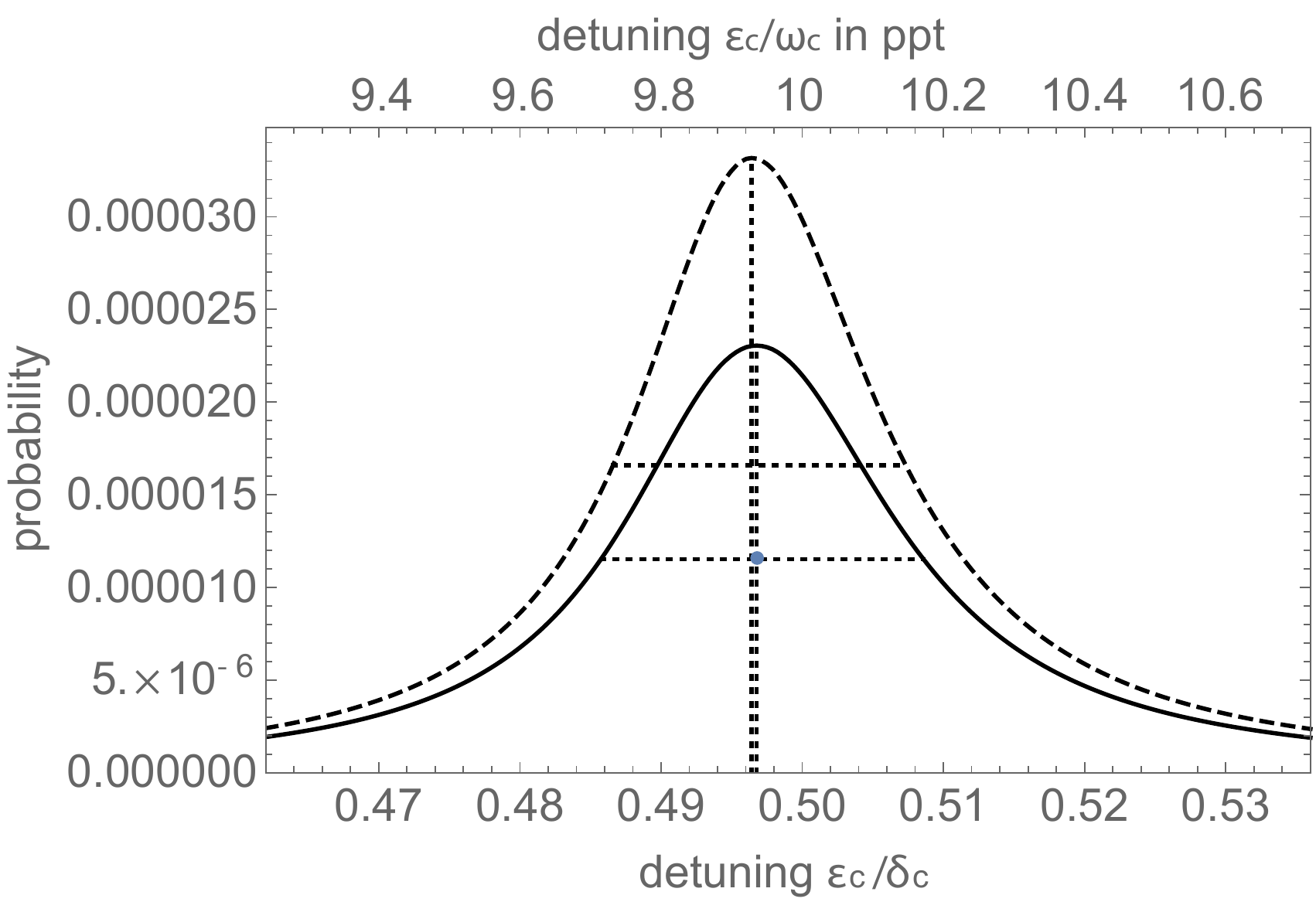}
 \caption{Cyclotron resonance line shape that is a horizontal slice through the maximum spin-flip probability of the contour plot in Fig.~\ref{fig:SimultaneousContour}(a).  The quasi steady state line shape (dashed) has the same shape as is numerically calculated (solid) but with a slightly different amplitude. 
    }    
    \label{fig:SimultaneousCyclotronLineshape}
\end{figure}

\begin{figure}[htbp!]
    \centering
    \includegraphics[width=\the\columnwidth]{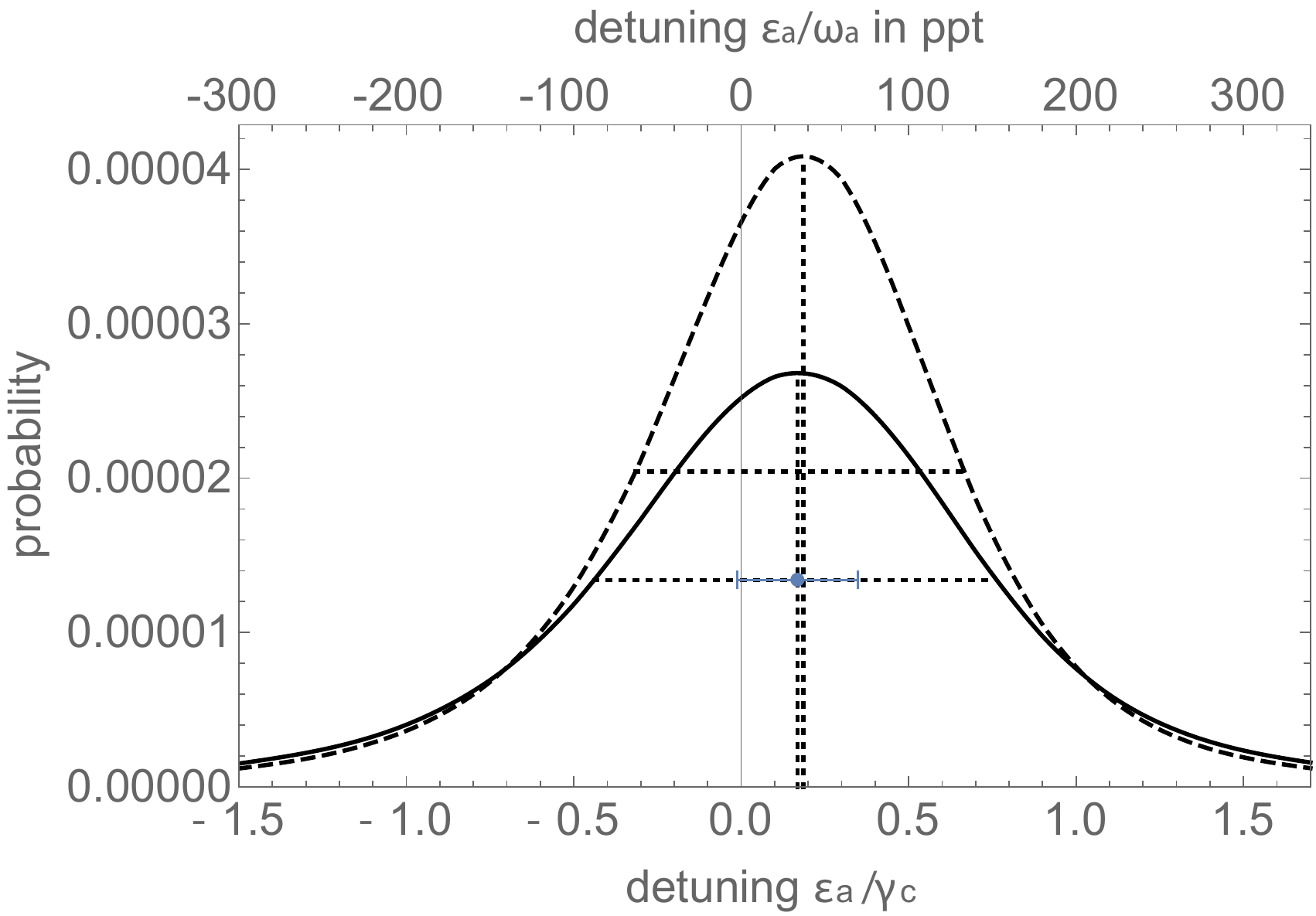}
    \caption{Anomaly line shape that is a vertical slice through the maximum spin-flip probability of the contour plot in Fig.~\ref{fig:SimultaneousContour}(a).  The quasi steady state line shape (dashed) has the same shape as is numerically calculated (solid) but with a slightly different amplitude. 
   }    
    \label{fig:SimultaneousAnomalyLineshape}
\end{figure}

A significant challenge to using two drives is the extremely small transition rates.  The maximum rate is for a cyclotron drive detuning $\epsilon_c = \delta_c/2$ (Fig.~\ref{fig:SimultaneousCyclotronLineshape}) and a anomaly detuning of about $\epsilon_a = 0.2\gamma_c=2\delta_a$ (Fig.~\ref{fig:SimultaneousAnomalyLineshape}).  Weak drives, with $\Omega_c = \Omega_a = \gamma_c/10$, applied for a time $t = 10 /\gamma_c \approx 53$ s, a time long enough for cyclotron transients to die out, avoids power broadening of the resonance lines. However, the transition probability in time $t$ is then approximately given by $\frac{d P_3}{dt} \times t$. The challenge is that the maximum transition probability is then about $2.5 \times 10^{-5}$ (Fig.~\ref{fig:SimultaneousD3}). This is a factor of $10^3$ times smaller than the peak cyclotron excitation rate for the one-drive case, and is likely too small to be useful.  A careful study will thus be required to determine the drive strengths and durations that can be used to get an acceptable rate and broadening. This seems possible, and such transitions have been used in experiments to prepare the desired spin state for measurement \cite{HarvardMagneticMoment2008}.  However, numerical solutions of the differential equations will be required since the weak drive limit will not apply.  The optimisation, when it is well motivated, would take some time to carry out given the size of the parameter space, even though we have demonstrated that it is feasible with the codes available.  

An intriguing possibility is that measurements made using simultaneous cyclotron and anomaly drives might be much less sensitive to magnetic field drifts because an cyclotron excitation and an anomaly transition would both take place before the field could drift much. (Considerable time passed between the measurements of these frequencies in past measurements.)  A study of this possibility would also require using the numerical solutions of the differential equations along with a realistic model of linear or quadratic magnetic field drift. Again, although we have demonstrated that this should be possible, it would take considerable time to carry this out.  

\section{Prospects for Electron and Positron Magnetic Moment Measurements with Significant Accuracy Improvements}
\label{sec:Comparison}

\subsection{First Possibility}

As discussed in Sec.~\ref{sec:QuantumCyclotron}, two extremely precise frequency measurements must be made to use the quantum cyclotron to determine the electron or positron much more precisely.  For all measurements so far, the anomaly and cyclotron frequencies have been measured, and 
\begin{equation}
\pm \mu/\mu_B = 1 + \omega_a/\omega_c    
\end{equation}
has been used to deduce the magnetic moment in Bohr magnetons. A $3 \times 10^{-14}$ measurement, ten times more precise than existing measurements, requires that the frequency ratio be measured to $3 \times 10^{-11}$. 

For a cyclotron frequency measurement at 150 GHz, the largest impediment 
to an improved measurement is the detection backaction width $\bar{n}_z \delta_c$ that is of order 30 Hz and $2 \times 10^{-10}$. Fortunately,  our proposal to circumvent detector back action \cite{Fan2020BackActionPRL} elaborated in this work provides a way to keep this backaction from contributing to the uncertainty of a new measurement. We showed that the remaining cyclotron line shape is very symmetric, with widths coming from cyclotron damping width $\gamma_c$ and axial decoherence $\bar{n}_z \gamma_z$. In Sec.~\ref{sec:OneDriveExcitations} we saw that these contribution together were about 3 time the cyclotron damping width, a width of 0.1 Hz and fractional width of $6 \times 10^{-13}$.  These values are a factor of 50 smaller than is needed for the the contemplated measurement.

With the cyclotron detection backaction circumvented, the anomaly frequency uncertainty becomes the largest challenge.  The three linewidth contributions are $\gamma_c$ from cyclotron decay, $\bar{n}_z \gamma_z$ from axial state decoherence, and $\bar{n_z}\delta_a$ from detection backaction, the latter contributing asymmetry to the line shape.  These are all comparable in size at about 0.03 Hz which, because the cyclotron frequency is 1000 times smaller than the anomaly frequency, is a much larger fractional uncertainty of about $2\times 10^{-10}$.  The desired measurement uncertainty thus seems attainable if the anomaly frequency can be extracted from the resonance line shape with an uncertainty ten times smaller than these contributions to the anomaly line shape.   This work thus suggests that a ten times improved measurement should be possible.

\subsection{Second Possibility}

Because the uncertainty in the anomaly frequency now seems to be the largest challenge for an improved measurement, we consider the option of instead determining the magnetic moment from the ratio of the spin and cyclotron frequencies,
\begin{equation}
    \pm \mu/\mu_B = \omega_s/\omega_c.  
\end{equation}
A direct spin-flip drive (Sec.~\ref{sec:DirectSpinFlips}) or simultaneous cyclotron and anomaly drives (Sec.~\ref{sec:twophotoncalculation}) to determine $\omega_s$.  The daunting challenge is that this frequency ratio must then be determined to the desired precision in the electron and positron magnetic moment of $3 \times 10^{14}$, a factor of 1000 better than for possibility one above.  

For the cyclotron linewidth of of 0.1 Hz and fractional linewidth of $6 \times 10^{-13}$ noted above, the cyclotron frequency would need to be extracted to a precision that was at least 30 times narrower than the anticipated linewidth.  This may not be an unreasonable linesplitting given that the line shape should be symmetric about the cyclotron frequency once the the detection backaction is circumvented. 

Directly driving spin flips to determine the spin frequency to the same precision would also be required, the first time that this would be realized with a quantum cyclotron. A two photon cyclotron plus anomaly transition would be an alternative. As for the cyclotron line shape, detection backaction that would make the resonance line shape broad and asymmetric can be circumvented. Because the two spin states are effectively stable, there would be no contributions to the line width from decay of an unstable state though the axial decoherence width $\bar{n}_z \gamma_z$ would persist.  This is an alternative route to a new measurement, in principle.   

\subsection{Magnetic Field Instability}

The quantum calculations support the viability of both of the measurement possibilities outlined above.  For the immediate future, however, measurements will almost certainly rely upon the first possibility -- measuring an electron or positron's  anomaly and cyclotron frequencies.  The reason is that the magnetic field produced by the best of superconducting solenoids drifts in time.  The demonstrated drift rates (about 1 part in $10^{10}$ per hour \cite{Helium3NMR2019}) is slow enough to make it possible to alternate determinations of the anomaly and cyclotron frequencies rapidly enough to make a new measurement.  This source of systematic uncertainty had to be carefully managed already in past measurements \cite{HarvardMagneticMoment2011}.  

To obtain the same precision using the second measurement possibility, alternating instead measurements of the spin and cyclotron frequencies, requires measuring these frequencies 1000 times more rapidly or producing a much more stable magnetic field.  The source of laboratory magnetic field instability and its reduction, whether by better solenoid design or shielding against changes in magnetic flux, is an interesting and important topic but it is beyond the scope of this calculation.

\section{Summary and Conclusions}
\label{sec:summary}

A quantum calculation is carried out for a driven one-electron quantum cyclotron with a quantum nondemolition (QND) coupling to a harmonic detection motion. The quantum spin and cyclotron motions have a QND coupling to a quantum axial detection motion, which in turn is coupled to a thermal reservoir.  External drives are applied to produce one-quantum transitions between the lowest spin and cyclotron states.  

A master equation is used to describe the driven motion of this open quantum system.  Convenient steady state solutions and resonance lineshapes for weak drives are presented, illustrated and discussed.  Numerical solutions reveal the time evolution and check the steady state line shapes.  Calculations of driven cyclotron excitations and driven anomaly transitions are presented, along with calculations for  directly driven spin flips and spin flips driven by simultaneous anomaly and cyclotron drives.  For a next generation of measurements, the first two of these four drive options turn out to be the most promising.  For weakly driven cyclotron and spin excitations, the predicted steady-state lineshapes for experimental parameters that have recently become accessible, are very different than the Brownian motion prediction used to interpret past measurements. 

An exciting result is the emergence of extremely narrow quantum resonances that appear within the cyclotron resonance line, corresponding to resolved quantum states of the axial detection oscillator.  These symmetric lines are about 100 time narrower than the broad and asymmetric cyclotron line shape that has been the biggest obstacle to a new generation of magnetic moment measurements.  Resolving these narrow peaks circumvents the detection backaction that would otherwise cause broad and asymmetric cyclotron resonance lines, reducing it to what is caused by only the zero-point motion of the detection motion, even many more detection states are populated.  The circumvention opens the way to the much more precise measurements of the cyclotron frequency that are needed to determine the electron and positron magnetic moments.  

Given the new method to measure the cyclotron frequency extremely accuracy, measuring the anomaly frequency precisely will become the biggest challenge to more precise magnetic moment measurements.  The anomaly line shape cannot be resolved into narrow symmetric peaks that correspond to individual quantum states of the axial detection motion. Nonetheless, the calculations suggest that an anomaly line shape can be produced that will make possible measurements that are perhaps an order of magnitude more precise. An initial survey of the effect of changing experimental parameters (e.g.\ cyclotron damping rate and lower ambient temperature) upon the anomaly lineshape identifies possible future upgrade paths, though none of these by itself is a large step.

The electron and positron magnetic moments are the most precise predictions of the Standard Model of Particle Physics -- the fundamental mathematical description of physical reality.  
Whether the current discrepancy between the measured electron magnetic moment and the Standard Model prediction is a hint of physics beyond the Standard Model 
is not yet known, but it warrants investigation.  The calculation and methods in this work indicate how this may be possible.

\bigskip
\section{acknowledgements}
 A preliminary version of some of this work is in a thesis \cite{ThesisDUrso}. This work was supported by the NSF, with X.\ Fan partially supported by the Masason Foundation. B.\ D'Urso, S.\ E.\ Fayer, T.\ G.\ Myers, B.\ A.\ D.\ Sukra and G.\ Nahal provided useful comments. 

\bibliographystyle{prsty_gg}
\bibliography{ggrefs2018,NewRefs}

\end{document}